\documentclass[a4paper,11pt]{article}
\usepackage{jheppub} 
\usepackage[T1]{fontenc}
\usepackage{physics}
\RequirePackage{verbatim}
\NewDocumentCommand{\tens}{t_}
{%
	\IfBooleanTF{#1}
	{\tensop}
	{\otimes}%
}
\NewDocumentCommand{\tensop}{m}
{%
	\mathbin{\mathop{\otimes}\displaylimits_{#1}}%
}
\usepackage{amsmath}
\usepackage{color}
\usepackage{amsfonts}
\usepackage{verbatim}
\usepackage[toc,page]{appendix}
\usepackage{amssymb}
\usepackage{bbold}
\usepackage{graphicx}
\usepackage{pgfplots}
\usepackage{float}
\title{Symmetries of $\kappa$-Minkowski space-time: A possibility of exotic momentum space  geometry?}
\author[a,b]{Partha Nandi,}
\author[1,b]{Anwesha Chakraborty,\note{Corresponding author}}
\author[c,b]{Sayan Kumar Pal,}
\author[d]{Biswajit Chakraborty,}
\author[a]{Frederik G Scholtz}
\affiliation[a]{Institute of Theoretical Physics, \\University 
	of Stellenbosch, Stellenbosch-7600, South Africa}
\affiliation[b]{Department of Astrophysics and High Energy Physics,\\S.N. Bose National Centre for Basic Sciences,\\Block-JD, Sector-III, Kolkata-700106, West Bengal, India}
\affiliation[c]{Department of Physics, Tezpur University,\\
	Tezpur, Assam-784028, India}
\affiliation[d]{Department of Physics, School of Mathematical Sciences,\\Ramakrishna Mission Vivekananda Educational and Research Institute,\\Belur Math, Howrah-711202,  West Bengal, India}

\emailAdd{pnandi@sun.ac.za} \emailAdd{anwesha@bose.res.in} \emailAdd{pal.sayan566@gmail.com} \emailAdd{dhrubashillong@gmail.com} \emailAdd{fgs@sun.ac.za}

\abstract{The quest for a quantum gravity phenomenology has inspired a quantum notion of space-time, which motivates us to study the fate of the relativistic symmetries of a  particular model of quantum space-time, as well as its intimate connection with the plausible emergent curved \textquotedblleft{physical} momentum space". We here focus on the problem of  Poincare symmetry of  $\kappa$-Minkowski type non-commutative (quantum) space-time, where the Poincare algebra, on its own, remains undeformed, but  in order to preserve the structure of the space-time non-commutative (NC) algebra, the actions of the algebra generators on the operator-valued space-time manifold must be enveloping algebra valued that lives in entire phase space i.e. the cotangent bundle on the space-time manifold (at classical level). Further, we constructed a model for a spin-less relativistic massive particle enjoying the deformed Poincare symmetry, using the first order form of geometric Lagrangian, that satisfies a new deformed dispersion relation and explored a feasible regime of a future Quantum Gravity theory in which the momentum space becomes curved. In this scenario there is only a mass scale (Planck mass $m_{p}$), but no length scale. Finally, we relate the deformed mass shell to the geodesic distance in this curved momentum space, where the mass of the particle gets renormalized as a result of noncommutativity. We show, that under some circumstances, the Planck mass provides an upper bound for the observed renormalized mass.\\\\
	Keywords: Symmetry, Quantum space-time, Relativistic free particle, Curved momentum space, Deformed dispersion relation}
\begin{document}
	\maketitle
	\flushbottom
	\section{Introduction}
	General Theory of Relativity (GTR), as formulated by Einstein more than a century back, is basically a classical field theory of gravitation. Its various predictions are being subjected to stringent experimental tests with ever increasing accuracy even now and that too in both the strong and weak gravity regimes. For example, we can mention the recent observations of images of shadows of supermassive blackholes (BH) and the detection of gravitational waves \cite{abbott} respectively, which has been verified with unprecedented accuracy. Despite all these successes, there remains serious conceptual shortcomings, like the occurrence of a singularity, as demonstrated, for example, through the BH singularity or the geodesically incomplete nature of space-time itself as demonstrated by the celebrated Hawking-Penrose singularity theorem \cite{penrose,elise}. In a  sense, therefore, GTR foretells its own demise and the classical theory is expected to be overtaken by some form of yet-to-be-developed theory of quantum gravity (QG) with the expected length scale where this happens being the Planck length $L_p=\sqrt{\hbar G} \sim 10^{-33}$ cm or equivalently
	a mass/energy scale $\kappa =\sqrt{\hbar/G} \sim 10^{19}$ GeV (we are using c=1). In fact, this has been the holy grail of theoretical physics for nearly a century because of the enormous success of quantum mechanics, like GTR.  Advances in current technologies have made it possible to access physical phenomena at much smaller length scales and one may expect to see new phenomena that cannot be accommodated in the existing theoretical framework.\\
	Besides this, there are strong plausibility arguments \cite{doplicher,seiberg}, suggesting a quantum (non-commutative) nature of space-time in the vicinity of these tiny length or large energy scales, where the space-time coordinates are promoted to the level of operators fulfilling a non-vanishing commutator algebra. Apart from 
	furnishing a natural mechanism to evade a gravitational collapse, that might occur in any attempt to localise an event down to this length scale, it paves the way to formulate a consistent coupling of gravity in the form of
	quantum space-time with the quantum fields occurring in the Standard model of particle physics, as the latter cannot be coupled to classical space-time \cite{vedral}.  This coupling should finally result in the emergence of GTR through some sort of coarse-graining procedure. Even a semi-classical approach will not serve its purpose, as BH evaporation through Hawking radiation, gives rise to the well known information paradox, as this is in conflict with unitary time evolution of quantum systems \cite{hawking}. Thus it becomes quite imperative for us to look for ways to formulate a full-fledged theory of QG. But, perhaps the biggest obstacle in this approach,
	remains the dearth of any experimental inputs, particularly in the high energy realms. Interestingly, it has been proposed quite recently in \cite{bose} that one can envisage a possible impact on the quantum origin of gravity even in the regime of weak Newtonian gravity through quantum superposition and quantum entanglement in the infrared region \cite{marletto}. See also \cite{marletto2} for a similar suggestion. It has some points of contact with the inherent entanglement, seen in the quantum gravity induced entanglement of masses (QGEM). This motivates us to look for systems exhibiting some plausible and robust features that may survive in an 
	appropriately chosen regime of a future theory of QG theory.  For this, we can perhaps consider a system in a regime, where the associated length scale $l>>L_p$, but the mass scale $m$ is comparable to $\kappa : m \leq \kappa$. With this,we 
	are effectively considering a scenario where both $\hbar$ and $G$ tend to zero,
	but their ratio is held fixed: $\sqrt{\hbar /G}=\kappa$. Further, the noncommutative coordinate algebra, for which the system dynamics can essentially be analysed
	classically, are those where the NC parameters have length dimensions and can be identified with $\sim L_p$ itself and they naturally occur in Lie-algebra type of noncommutativity, as the deformation parameter. This will effectively reduce to $\kappa^{-1}$ in the associated Poisson (or Dirac)
	bracket in the corresponding classical description. For that we can naturally
	consider the well known $\kappa$-Minkowski spacetime \cite{luk2,luk1,majid3,luk4,lukierski}, which incidentally was proposed by \cite{camelia/dsr,camelia/dsr2,glikmannew} in the context of double-special relativity, where attempts were
	made to deform the special theory of relativity (STR) further by incorporating this 
	new observer independent scale $\kappa$. The presence of this mass scale $\kappa$ alone can, however, have a drastic 
	impact on the system dynamics, like curving the energy-momentum or simply the momentum space \cite{smolin}, thereby deforming the dispersion relation. Furthermore, a curved momentum space may force us to forgo absolute locality and to embrace \textquotedblleft{relative locality}" in its place \cite{smolin,freidel1}. Incidentally, it was Max Born, who had speculated about the necessity of curved momentum space in the context of QG way back in 1938 \cite{born}. Later in the year 2000, it was shown again in \cite{camelia-shahn,majid} that the curved momentum space can be thought of as Hopf dual to NC space-time and the term `co-gravity' was coined by the authors in this context. This is reminiscent of the well known curving of 3-velocity space
	in STR, where the deformation parameter is the speed of light $c$. This is manifested through the non-linear addition of velocities, which is generically noncommutative  and non-associative in nature. Furthermore, like in STR, here too the flat limit is recovered only in the limit $\kappa \rightarrow \infty$. A concrete realization of this through a model displaying curved momentum space was provided in a $2+1$-dimensional system, where a spin-less relativistic point particle interacts with Einstein's gravity. Here the latter is a topological theory and can be cast in the form of a non-abelian ($ISO(2,1)$) Chern-Simons theory \cite{witten}. There is some indirect evidence for this being true even for the realistic $3+1$-dimensional spacetime, although Einstein's theory is no-longer topological in nature \cite{simone}. However, it was shown in \cite{glikman1} that a $\kappa$-Poincare-invariant action for a relativistic particle moving in a restricted form of $\kappa$-Minkowski spacetime can indeed be constructed by using certain group theoretical formulations, where the momentum space can be identified with the group manifold $AN(3)$ \cite{glikman2}, which is a part of the de Sitter (dS) spacetime and the coordinate operators are the corresponding generators of translation. In this construction the translation generators $p^{\mu}$'s do not transform as Lorentz 4-vectors, rather the transformation of $p^{\mu}$ is given by an element of the enveloping  algebra. This motivates us to consider a more general kind of $\kappa$-Minkowski spacetime, where both the action and the entire set of commutator algebra respects the \textit{undeformed} Poincare symmetry. We find that it becomes essential to deform the actions of the Poincare generators on the coordinate operators, so that they are now allowed to take values in the universal enveloping algebra, obtained from the
	undeformed Poincare generators, but in a manner that the Poincare algebra-by itself undergoes no
	deformation. In this context, we find it convenient to follow the template of \cite{koch},
	where the symmetries of Moyal space-time has been discussed. Also, we could not adopt the twisted Hopf-algebraic framework, as no twist or star product seems to exist in this
	case, although there exist other variants of $\kappa$-Minkowski spacetime \cite{lukierski,jerzy,juric,twist,meljanac2,woronowicz}, where both the star product/twist can be readily introduced.\\\\
	Importantly, we carry out the analysis entirely at the classical level by demoting
	all the generators to commuting classical variables, where the various commutator
	brackets now correspond to usual symplectic brackets $\{ . , . \}$. We then show how these brackets can, in turn, be interpreted as the Dirac bracket of a first order constrained system, describing the dynamics of the relativistic particles moving
	in the $\kappa$-Minkowski spacetime. We find that 
	$\{x^i, p_j\}$ undergoes a momentum dependent deformation, so that the momentum space can now be identified with curved space, which, however, is not quite Riemannian. Nevertheless, an invariant line element of the form $ds= \sqrt{g_{\mu \nu}dp^{\mu}dp^{\nu}}$ can be introduced, so that the geodesic distance in the momentum space can easily be computed, enabling us to obtain  the deformed dispersion relation which eventually helps us to identify renormalised observable mass $M$. And for any isolated fundamental particle the mass scale $\kappa$  can serve as an upper bound for this renormalized mass : $M  < \kappa$ for a certain choice of the noncommutative parameter. Interestingly, it turns out that this $\kappa$ is the only surviving natural mass scale in this classical system.\\\\
	The paper is organized as follows: In section-2 we revisit the deformed symmetries of $\kappa$-Minkowski space-time using the formulation \cite{koch} followed by a discussion on the deformed co-products of Poincare generators and a Heisenberg-double construction in Hopf-algebroid framework for a consistency check of the phase-space algebra coming from the deformed co-product of the momentum. In section-3 we construct the first order Lagrangian of a free massive relativistic spinless particle moving in $\kappa$ space-time which obeys the same symmetry as that of the space-time. In subsection-3.1 we explicitly derive the deformed mass-shell condition using the computation of geodesic distance in momentum space. In sub-section-3.2, we have shown the non-canonical transformation between non-commutative and commutative coordinates, which helps us to find out the explicit form of the deformed Lorentz generators. Later in sub-section-3.3 we show the invariance of the Lagrangian under the deformed symmetries and derived the Poincare generators using Noether's prescription and confirmed them with the ones derived in the preceding sub-section. In sub-section-3.4 we have studied the feasibility of lifting the infinitesimal Lorentz symmetry to a finite one on the NC coordinate, using the non-canonical transformation between commutative and NC coordinates derived earlier in sub-section-3.2 and found an explicit Lorentz invariant 'interval' under the finite transformation which is now a function of phase-space variables. Finally, in section-4 we conclude with some overview and future directions.
	\section{Deformed symmetries of $\kappa$ Minkowski spacetime}
	The $\kappa$ Minkowski space-time $\mathcal{\hat{M}}$ is introduced through the coordinate algebra  given by  
	\begin{equation}
		[\hat{X}^{\mu},\hat{X}^{\nu}]=i\hat{\theta}^{\mu\nu}=i(a^{\mu}\hat{X}^{\nu}-a^{\nu}\hat{X}^{\mu})\label{1}
	\end{equation}
	where $\hat{X}^{\mu} \in \mathcal{\hat{M}}$ are the operator valued noncommutative coordinates and $a^{\mu}$ - a set of four constants, which are real scalars and can be identified with the set of four deformation parameters \footnote{The deformation parameters are of the order of $\kappa^{-1} : a^{\mu} \sim \kappa^{-1}$.}. Note that despite of its appearance $a^{\mu}$ does not transform like a vector under Lorentz transformation, rather all its components remain the same in every Lorentz frame\footnote{By this we mean that $a^{\mu}$'s commute with all the Poincare generators : $[M_{\nu\lambda},a^{\mu}]=[\hat{P}_{\nu},a^{\mu}]=0$. This, however, does not prevent us from introducing $a_0=a^0$ and $a_i=-a^i (i=1,2,3)$ as another set of scalars. And here too, we can raise/lower indices  using $\eta_{\mu\nu}$ and write formally $a_{\mu}=\eta_{\mu\nu}a^{\nu}$ and we can write $a^{\mu}a_{\mu}= a_0^2-\vec{a}^2$. Further depending on whether $a^{\mu}a_{\mu}> 0 
		, < 0, =0 $ we can refer to it as time-like, space-like and null respectively \cite{dimitrijevic}.}. It can be checked easily that under usual infinitesimal translation and homogeneous Lorentz transformation of coordinates given by
	\begin{align}
		& \delta 
		\hat{X}^{\mu}=\epsilon^{\mu};\nonumber\\
		& \delta\hat{X}^{\mu}= \omega^{\mu}\,_{\alpha}\hat{X}^{\alpha}\label{a1}
	\end{align}
	(	where $\epsilon^{\mu}$ and $\omega^{\mu\nu}=-\omega^{\nu\mu}$ with $|\epsilon^{\mu}|, |\omega^{\mu\nu}| << 1$ are infinitesimal parameters corresponding to translations and homogeneous Lorentz transformations), 
	the coordinate algebra (\ref{1}) does not retain its primitive form. This indicates that some sort of a deformed Poincare transformation is required to preserve the $\kappa$ -Minkowski algebra (\ref{1}), under the respective deformed transformations. In this section we shall derive the deformed symmetries corresponding to Poincare transformation, which we will achieve without deforming the $\mathfrak{iso}$(1,3) Lie algebra between the generators of Lorentz transformation and translation; only their actions on the coordinate operators will be deformed. This ensures the vector like transformations of the translation generators under Lorentz transformation\footnote{One may deform even the $\mathfrak{iso}$(1,3) Lie algebra itself to obtain $\kappa$-Poincare algebra \cite{majid3,glikman}, which too can generate deformed transformation of the coordinates, keeping the algebra (\ref{1}) stable under such transformation \cite{dimitrijevic}. However in that case, the translation generator will not transform vectorially under Lorentz transformation; rather they will give enveloping algebra valued transformation.}. Although this has been shown by various authors in the literature \cite{meljanac2,juric,juric2,samsarov,dimitrijevic,dimitrijevic2,meljanac1}, we revisit the problem here again to provide a systematic derivation for obtaining the deformed transformations, following the template given by F. Koch \textit{et.al.} \cite{koch}. As we shall see, this exercise becomes important for our discussion in the subsequent sections.
	Particularly this will help us to construct a relativistic action of a free spin-less point particle with non-vanishing mass, moving in $\kappa$ Minkowski space-time $\mathcal{\hat{M}}$. \\
	The usual $\mathfrak{iso}$(1,3) Lie algebra between the Poincare generators, is given by,  
	\begin{align}
		[\hat{M}_{\mu\nu},\hat{M}_{\rho\sigma}]&=i(\eta_{\nu\rho}\hat{M}_{\mu\sigma}+\eta_{\mu\sigma}\hat{M}_{\nu\rho}-\eta_{\mu\rho}\hat{M}_{\nu\sigma}-\eta_{\nu\sigma}\hat{M}_{\mu\rho})\nonumber\\
		[\hat{M}_{\mu\nu},\hat{P}_{\rho}]&=i(\eta_{\nu\rho}\hat{P}_{\mu}-\eta_{\mu\rho}\hat{P}_{\nu})\nonumber\\
		[\hat{P}_{\mu},\hat{P}_{\nu}]&=0\label{2}
	\end{align}
	where $\hat{M}_{\mu\nu}$ and $\hat{P}_{\mu}$ refers to  Lorentz and translation generators respectively giving in total 10 generators of the Poincare algebra. As mentioned above our goal is to find the deformed transformations in the coordinate sector, without deforming the Lie algebra (\ref{2}) between the Poincare generators themselves. It is therefore quite clear that the transformation of $\hat{P}_{\mu}$ under Poincare transformation remains unchanged, giving us its infinitesimal transformation corresponding to translation and Lorentz transformation as 
	\begin{equation}
		\delta \hat{P}_{\mu}=0 \qquad\textrm{and}\quad \delta\hat{P}_{\mu}=\omega_{\mu}\,^{\alpha}\hat{P}_{\alpha}
	\end{equation} 
	respectively.\\
	It is speculated that, to keep the coordinate algebra (\ref{1}) consistent under the action of the Poincare generators, satisfying the above Lie algebra (\ref{2}), one must deform the infinitesimal transformations (\ref{a1}) thereby deforming the commutators $[\hat{M}_{\mu\nu},\hat{X}_{\rho}]$ and $[\hat{P}_{\mu},\hat{X}_{\nu}]$. With that in mind, let us make the following ansatz for the deformed brackets as, 
	\begin{align}
		[\hat{M}_{\mu\nu},\hat{X}_{\rho}]&=i(\eta_{\nu\rho}\hat{X}_{\mu}-\eta_{\mu\rho}\hat{X}_{\nu})+i\psi_{\mu\nu\rho}(\hat{P},\hat{M})\label{3}\\
		[\hat{P}_{\mu},\hat{X}_{\nu}]&=-i\eta_{\mu\nu}\phi(\hat{P})+i\chi_{\mu\nu}(\hat{P},\hat{M})\label{4}
	\end{align}
	Note that the deformations are entirely contained in $\psi_{\mu\nu\rho},\phi,\chi_{\mu\nu}$ which we would like to depend entirely on the noncommutative parameter $a_{\mu}$ linearly and also Lie algebra valued i.e. depending on $\hat{M}$ and $\hat{P}$ linearly, rather than taking values in the universal enveloping algebra, so as to enable us to look for a simple and almost unique solution \cite{dimitrijevic}. However, as it turns out that this demand can't be complied with entirely and deformations involving higher orders of $a_{\mu}$ and valued in universal enveloping Lie algebra, must be invoked in a special case, particularly in the deformed Heisenberg algebra, in order to get consistent solutions (This point will be discussed thoroughly later in this section). We also require,  $\psi_{\mu\nu\rho}(\hat{P},\hat{M};a)\to 0,\,\phi(\hat{P};a)\to 1$ and $\chi_{\mu\nu}(\hat{P},\hat{M};a)\to 0$ for $a_{\mu} \to 0$, so that in the commutative limit the  brackets (\ref{3},\ref{4}) gives the usual results. Let us also point out that the particular dependence of the deformations $\psi$ and $\phi,\chi$ on the generators $\hat{P}$ or $\hat{M}$ are to be taken in such a way that their dimensions match with the left hand side of (\ref{3}) and (\ref{4}) respectively.\\\\
	In order to solve $\psi_{\mu\nu\rho}(\hat{P},\hat{M};a),  \phi(\hat{P};a),  \chi_{\mu\nu}(\hat{P},\hat{M};a)$ , we invoke two kinds of following consistency conditions \cite{lemos}:\\ \\
	\textbf{Condition-1:}   If we consider the infinitesimal Poincare transformation:
	\begin{equation}
		\hat{X}^{\mu}\rightarrow \hat{X}'^{\mu}=\hat{X}^{\mu}+\delta_{\epsilon} \hat{X}^{\mu},
	\end{equation}
	with $\delta_{\epsilon} \hat{X}^{\mu}=i \epsilon^{i}  [\hat{G}_{i},\hat{X}^{\mu}]$, where  generators for the respective infinitesimal transformation is indicated collectively by the symbol $\hat{G}_i \in \{\hat{M},\hat{P}\}$, then it may be noted that the covariance of the NC relations (\ref{1}) under such space time transformation leads to the Jacobi identities between $\hat{G}_{i}$, and two  non-commutative space time coordinates:
	
	\begin{equation}
		[[\hat{X}_{\mu},\hat{X}_{\nu}],\hat{G}_{i}]+[[\hat{X}_{\nu},\hat{G}_{i}],\hat{X}_{\mu}]+[[\hat{G}_{i},\hat{X}_{\mu}],\hat{X}_{\nu}]=0\label{10}
	\end{equation}
	Using various forms of $\hat{G}_{i}$, condition (\ref{10}) gives the following relations,
	\begin{align}
		[[\hat{P}_{\lambda},\hat{X}_{\mu}],\hat{X}_{\nu}]+[[\hat{X}_{\mu},\hat{X}_{\nu}],\hat{P}_{\lambda}]+[[\hat{X}_{\nu},\hat{P}_{\lambda}],\hat{X}_{\mu}]&=0\nonumber\\
		[[\hat{M}_{\rho\sigma},\hat{X}_{\mu}],\hat{X}_{\nu}]+[[\hat{X}_{\mu},\hat{X}_{\nu}],\hat{M}_{\rho\sigma}]+[[\hat{X}_{\nu},\hat{M}_{\rho\sigma}],\hat{X}_{\mu}]&=0\label{11}
	\end{align}
	Again inserting the commutators from (\ref{1},\ref{3},\ref{4}) we get a set of relation between  $\psi,\chi$ and $\phi$ as
	\small{
		\begin{align}
			&i\Big\{\eta_{\lambda\nu}[\phi(\hat{P}),\hat{X}_{\mu}]-\eta_{\lambda\mu}[\phi(\hat{P}),\hat{X}_{\nu}]\Big\}+i\Big\{[\chi_{\lambda\mu},\hat{X}_{\nu}]-[\chi_{\lambda\nu},\hat{X}_{\mu}]\Big\}+\Big(a_{\nu}\eta_{\lambda\mu}-a_{\mu}\eta_{\lambda\nu}\Big)\phi(\hat{P})+\Big(a_{\mu}\chi_{\lambda\nu}-a_{\nu}\chi_{\lambda\mu}\Big)=0\label{12}\\
			&i\Big\{[\psi_{\rho\sigma\mu},\hat{X}_{\nu}]-[\psi_{\rho\sigma\nu},\hat{X}_{\mu}]\Big\}+\Big(a_{\mu}\psi_{\rho\sigma\nu}-a_{\nu}\psi_{\rho\sigma\mu}\Big)+\Big(\eta_{\rho\mu}a_{\sigma}-\eta_{\sigma\mu}a_{\rho}\Big)\hat{X}_{\nu}+\Big(\eta_{\sigma\nu}a_{\rho}-\eta_{\rho\nu}a_{\sigma}\Big)\hat{X}_{\mu}=0\label{13}
	\end{align}}\\
	\vspace{0.5cm}
	\normalsize
	\textbf{Condition-2:} \\
	Covariance of the relations given in (\ref{3},\ref{4}) gives the Jacobi identities between two  $\mathfrak{iso}$(3,1) generators $\hat{G}_i=\{\hat{M}_{\mu\nu},\hat{P}_{\lambda}\}$ and coordinate $\hat{X}^{\mu}$'s :
	\begin{equation}
		[[\hat{G}_i,\hat{G}_j],\hat{X}_{\mu}]+[[\hat{G}_j,\hat{X}_{\mu}],\hat{G}_i]+[[\hat{X}_{\mu},\hat{G}_i],\hat{G}_j]=0\label{5}
	\end{equation}
	Putting various combinations of Poincare generators and coordinate in (\ref{5}) we get the following conditions:
	\begin{align}
		[[\hat{P}_{\mu},\hat{P}_{\nu}],\hat{X}_{\lambda}]+[[\hat{P}_{\nu},\hat{X}_{\lambda}],\hat{P}_{\mu}]+[[\hat{X}_{\lambda},\hat{P}_{\mu}],\hat{P}_{\nu}]=0\nonumber\\
		[[\hat{M}_{\mu\nu},\hat{M}_{\sigma\rho}],\hat{X}_{\lambda}]+[[\hat{M}_{\sigma\rho},\hat{X}_{\lambda}],\hat{M}_{\mu\nu}]+[[\hat{X}_{\lambda},\hat{M}_{\mu\nu}],\hat{M}_{\sigma\rho}]=0\nonumber\\
		[[\hat{M}_{\mu\nu},\hat{P}_{\sigma}],\hat{X}_{\lambda}]+[[\hat{P}_{\sigma},\hat{X}_{\lambda}],\hat{M}_{\mu\nu}]+[[\hat{X}_{\lambda},\hat{M}_{\mu\nu}],\hat{P}_{\sigma}]=0\label{6}
	\end{align}
	Inserting the commutator relations (\ref{2}, \ref{3}, \ref{4}) in the Jacobi identities (\ref{6}), we get following relations between the deformation functions $\psi,\phi$ and $\chi$:
	\begin{align}
		[\chi_{\nu\lambda}(\hat{P},\hat{M}),\hat{P}_{\mu}]-[\chi_{\mu\lambda}(\hat{P},\hat{M}),\hat{P}_{\nu}]&=0\label{7}\\
		i[\psi_{\sigma\rho\lambda},\hat{M}_{\mu\nu}]-i[\psi_{\mu\nu\lambda},\hat{M}_{\sigma\rho}]+\eta_{\mu\lambda}\psi_{\sigma\rho\nu}+\eta_{\rho\lambda}\psi_{\mu\nu\sigma}+\eta_{\mu\sigma}\psi_{\nu\rho\lambda}+\eta_{\nu\rho}\psi_{\mu\sigma\lambda}&\nonumber\\-\eta_{\nu\sigma}\psi_{\mu\rho\lambda}-\eta_{\mu\rho}\psi_{\nu\sigma\lambda}-\eta_{\nu\lambda}\psi_{\sigma\rho\mu}-\eta_{\sigma\lambda}\psi_{\mu\nu\rho}&=0\label{8}\\
		-i\eta_{\sigma\lambda}[\phi(\hat{P}),\hat{M}_{\mu\nu}]+i[\chi_{\sigma\lambda},\hat{M}_{\mu\nu}]-i[\psi_{\mu\nu\lambda},\hat{P}_{\sigma}]+(\eta_{\mu\sigma}\chi_{\nu\lambda}
		-\eta_{\nu\sigma}\chi_{\mu\lambda}+\eta_{\mu\lambda}\chi_{\sigma\nu}-\eta_{\nu\lambda}\chi_{\sigma\mu})&=0\label{9}
	\end{align}	
	At this point we make a suitable (as discussed above condition-1) ansatz for $\chi_{\mu\nu}(\hat{M},\hat{P}),\phi(\hat{P})$ and $\psi_{\mu\nu\lambda}(\hat{M},\hat{P})$  and put them back in (\ref{12},\ref{13},\ref{7},\ref{8},\ref{9}) to exactly solve them.\\
	From (\ref{3}), note that $\psi_{\mu\nu\lambda}(\hat{M},\hat{P})$ has dimension of length and is anti-symmetric in the index $\mu,\nu$. So we accordingly choose the ansatz of $\psi$, which is also first order in the noncommutative parameter $a_{\mu}$, as follows,
	\begin{equation}
		\psi_{\mu\nu\lambda}= s_1a_{\lambda}\hat{M}_{\mu\nu}+t_1(a_{\mu}\hat{M}_{\nu\lambda}-a_{\nu}\hat{M}_{\mu\lambda})+u_1a^{\rho}(\eta_{\nu\lambda}\hat{M}_{\rho\mu}-\eta_{\mu\lambda}\hat{M}_{\rho\nu})\label{14}
	\end{equation}
	where $s_1,t_1,u_1$ are dimensionless parameters. We would like to point out here that, inclusion of any higher order terms of $\hat{M}$ or $\hat{P}$ in the above ansatz, would also correspondingly increase the order of the deformation parameter $a_{\mu}$ to retain the dimension of $\psi_{\mu\nu\lambda}$ as that of length. This would have jeopardised our objective to keep terms in linear order of $a^{\mu}$.  Thus (\ref{14}) is the most general form of the ansatz for $\psi_{\mu\nu\lambda}$ in this scenario. After substituting the ansatz back in (\ref{13},\ref{8}), one can carry out a straightforward but lengthy calculation to find the following values of the parameters:
	$$t_1=-1;\qquad s_1=u_1=0$$
	Finally we arrive at the following form of $\psi_{\mu\nu\lambda}(\hat{P},\hat{M})$ as,
	\begin{equation}
		\psi_{\mu\nu\lambda}=-(a_{\mu}\hat{M}_{\nu\lambda}-a_{\nu}\hat{M}_{\mu\lambda})\label{15}
	\end{equation}
	Next we consider the ansatz for $\phi(\hat{P})$ and $\chi_{\mu\nu}(\hat{P},\hat{M})$. From (\ref{4}), it can be seen that $\chi_{\mu\nu}$ is dimensionless and hermitian and has no particular symmetry in its indices and should vanish in the commutative limit $a_{\mu} \to 0$. In addition, $\phi(\hat{P})$, being a  \textit{formal} scalar and dimensionless, can be a function of  $a^{\mu}\hat{P}_{\mu}$ and/or $a^2\hat{P}_{\mu}\hat{P}^{\mu}$ only. The latter functional dependence opens up the possibility of considering higher orders in the deformation parameters, which is in fact forced on us here to get a consistent solution. We therefore make the following ansatz:
	\begin{equation}
		\phi(\hat{P})=s_2a^{\mu}\hat{P}_{\mu}+(1+t_2a^2\hat{P}^2)^n;\,\,\,\,\chi_{\mu\nu}=s_3a_{\mu}\hat{P}_{\nu}+u_3a_{\nu}\hat{P}_{\mu}+t_3a^{\lambda}(\hat{P}_{\lambda}\hat{M}_{\mu\nu}+\hat{M}_{\mu\nu}\hat{P}_{\lambda})\label{16}
	\end{equation}
	Note that, $\phi(P) \to 1$ and $\chi_{\mu\nu} \to 0$ for $a^{\mu} \to 0$, which is required in order to produce proper commutative limit. Here $s_2,t_2,n,s_3,u_3,t_3$ are dimensionless parameters. Now putting $\chi_{\mu\nu}$ back in (\ref{7}), we get $t_3=0$, giving us $\chi_{\mu\nu}=s_3a_{\mu}\hat{P}_{\nu}+u_3a_{\nu}\hat{P}_{\mu}$. \\
	Substituting the ansatz of $\phi(\hat{P})$ and $\chi_{\mu\nu}$ in (\ref{9}), we get
	$$s_2=1,\,\,\, s_3=1,\,\,\,u_3=0.$$
	Thus $\chi_{\mu\nu}$ is now exactly solved and the ansatz for $\phi(\hat{P})$ reduces to: 
	\begin{equation}
		\chi_{\mu\nu}= a_{\mu}\hat{P}_{\nu} ;\qquad \phi(\hat{P})= a.\hat{P}+(1+t_2a^2\hat{P}^2)^n   \label{17}
	\end{equation}
	Finally, substituting $\phi(\hat{P})$ and $\chi_{\mu\nu}$ (\ref{17}), in (\ref{12}), we get $$n=\frac{1}{2} \,\,\, \textrm{and}\,\,\,  t_2=1.$$
	So eventually we have determined the deformed brackets which will help us to derive the deformed transformation under Poincare generators, as
	\begin{align}
		[\hat{M}_{\mu\nu},\hat{X}_{\rho}]&=i(\eta_{\nu\rho}\hat{X}_{\mu}-\eta_{\mu\rho}\hat{X}_{\nu})-i(a_{\mu}\hat{M}_{\nu\rho}-a_{\nu}\hat{M}_{\mu\rho})\label{a3}\\
		[\hat{P}_{\mu},\hat{X}_{\nu}]&=-i\eta_{\mu\nu}\phi(\hat{P})+ia_{\mu}\hat{P}_{\nu};\,\,\,\,\,\,\,\phi(\hat{P})=a^{\mu}\hat{P}_{\mu}+\sqrt{1+a^2\hat{P}^2}\label{19}
	\end{align}
	With these we reproduce the results which are already existing in the literature \cite{juric,wohlgenannt,dimitrijevic} etc. 
	One may check that, in the commutative limit $a^{\mu} \to 0$, the above deformed commutators indeed reproduce the usual brackets i.e.
	\begin{equation}
		[\hat{M}_{\mu\nu},\hat{X}_{\rho}]=i(\eta_{\nu\rho}\hat{X}_{\mu}-\eta_{\mu\rho}\hat{X}_{\nu});\qquad  	[\hat{P}_{\mu},\hat{X}_{\nu}]=-i\eta_{\mu\nu} 
	\end{equation}
	The deformed transformations compatible with the space-time algebra (\ref{1}) can now be written as 
	\begin{align}
		\textrm{Under deformed translation:} \,\,\,\delta\hat{X}^{\mu}&= i\epsilon^{\alpha}[\hat{P}_{\alpha},\hat{X}^{\mu}]= \epsilon^{\mu}\phi(\hat{P})-(\epsilon_{\nu}a^{\nu})\hat{P^{\mu}}\label{a4a}\\
		\textrm{Under deformed Lorentz transformation:}\,\,\,\delta \hat{X}^{\mu}&=\frac{i}{2}\omega^{\alpha\beta}[\hat{M}_{\alpha\beta},\hat{X}^{\mu}]= \omega^{\mu}\,_{\alpha}\hat{X}^{\alpha}+\omega^{\alpha\beta}a_{\alpha}\hat{M}_{\beta}\,^{\mu}\label{a4}
	\end{align}
	They reduce to the usual transformations given in (\ref{a1}) in the commutative limit.\\ 
	As we can see from (\ref{a4a},\ref{a4}), the coordinate operators $X^{\mu}$'s shows non-vector like transformations under the deformed Poincare transformations. One can easily recheck that these transformations preserves the structures of the coordinate algebra (\ref{1}). In other words, the structures of (\ref{1}) are stable under (\ref{a4a},\ref{a4}).\\
	It is to be noted that although the Poincare algebra is not deformed, the action of the Poincare generators on the module i.e. the algebra generated by the coordinates $\hat{X}^{\mu}$ is highly deformed due to the deformed nature of the space. In other words, the nature of the deformation is not reflected through the undeformed nature of the $\mathfrak{iso}(1,3)$ Lie algebra. Rather, as we shall demonstrate in the sequel, albeit at the classical level that this deformed action owes its origin to the deformation in the structure of the Lorentz generators $\hat{M}_{\mu\nu}$'s themselves, when expressed in terms of $\hat{X}_{\mu}$ and $\hat{P}_{\mu}$.
	\subsection{Deformed co-algebra and the construction of Heisenberg double}
	The symmetry underlying the $\kappa$-Minkowski space-time $\mathcal{\hat{M}}$ is captured through a deformed actions (\ref{a4a},\ref{a4}) and they stem from the associated deformed structure of the Lorentz generators $\hat{M}_{\mu\nu}$ themselves. But the algebraic sector of this deformed symmetry is the same as that of the undeformed Poincare algebra. A natural question then arises: how does the deformed action impact on the single and/or multiparticle dynamics. In fact, there is a considerable body of literature \cite{juric,dimitrijevic,dimitrijevic2,meljanac1,kovacevic}, where the authors have already calculated the deformed co-algebra structures i.e. the deformed coproducts  ($\Delta$), deformed antipodes ($S$) and co-units ($\epsilon$). Although, these co-algebraic structures have no direct relevance for our study of kinematics and dynamics of  a single particle system, it nevertheless can be used to provide an alternative `derivation' of the deformed Heisenberg algebra (\ref{19}) through the construction of Heisenberg double in the framework of Hopf algebroid. For this, we shall essentially be following \cite{lukierski,meljanac1}. So to begin with, let us quote the expressions of the coalgebraic structures from the literature.
	\begin{align}
		\Delta (\hat{P}_{\mu})= &\hat{P}_{\mu}\otimes\phi+\textbf{1}\otimes \hat{P}_{\mu}-a_{\mu}(\hat{P}_{\nu}\phi^{-1})\otimes P^{\nu}+\frac{a_{\mu}}{2}(F(P)\phi^{-1})\otimes (a.\hat{P})\label{w1}\\
		\Delta(\hat{M}_{\mu\nu})= &\hat{M}_{\mu\nu}\otimes\textbf{1}+\textbf{1}\otimes \hat{M}_{\mu\nu}+ a_{\mu}\Big(\hat{P}^{\lambda}-\frac{a^{\lambda}}{2}F(\hat{P})\Big)\phi^{-1}\otimes \hat{M}_{\lambda\nu}- \nonumber\\
		&a_{\nu}\Big(\hat{P}^{\lambda}-\frac{a^{\lambda}}{2}F(\hat{P})\Big)\phi^{-1}\otimes \hat{M}_{\lambda\mu}\label{n2}\\
		\textrm{where}\qquad\quad&\nonumber\\
		F(\hat{P})&=\frac{2}{a^2}\Big(1-\sqrt{1+a^2\hat{P}^2} \Big)\nonumber
	\end{align}
	We have provided a brief derivations of these coproducts in the Appendix-\ref{app2a} \footnote{Actually, it will become clear in the sequel that this exercise is tantamount to a verification of the self-consistency of the deformed Heisenberg algebra \eqref{19}. This is because, \eqref{19} is used in Appendix-\ref{app2a} to derive the coproduct $\Delta(P_{\mu}), \Delta(M_{\mu\nu})$ \eqref{w1},\eqref{n2}. On the other hand, these coproducts are used in this sub-section to derive \eqref{19}.}. And the deformed antipodes $S(\hat{M}_{\mu\nu})$ and $S(\hat{P}_\mu)$ are given by
	\begin{align}
		S(\hat{P}_{\mu}) &= \Big(-\hat{P}_{\mu} + ia_{\mu} (\hat{P}_{\alpha}-\frac{ia_{\alpha}}{2}F(\hat{P}))\hat{P}^{\alpha}\Big)\phi^{-1}\\
		S(\hat{M}_{\mu\nu}) &= -\hat{M}_{\mu\nu} + ia_{\mu}\Big(\hat{P}_{\alpha}-\frac{ia_{\alpha}}{2}F(\hat{P})\Big)\hat{M}_{\alpha\nu}-ia_{\nu}\Big(\hat{P}_{\alpha}-\frac{ia_{\alpha}}{2}F(\hat{P})\Big)\hat{M}_{\alpha\mu}\label{n4}
	\end{align}
	fulfilling,
	$$m\Big[(\textbf{1}\otimes S)\Delta\Big]= m\Big[(S\otimes\textbf{1})\Delta\Big]= \eta \circ \epsilon$$
	where $\eta$ is unit of the algebra and $\epsilon$ is the co-unit given by 
	\begin{equation}
		\epsilon(\hat{P}_{\mu})=\epsilon(\hat{M}_{\mu\nu})=0\label{j5}
	\end{equation}
	which remains undeformed. One can observe at this stage that appropriate commutative limits i.e. the so called primitive forms of these co-algebraic structures are easily obtained in limit $a_{\mu}\to 0$.
	\begin{align}    \Delta(\hat{P}_{\mu})&\to\Delta_0(\hat{P}_{\mu})=\hat{P}_{\mu}\otimes\textbf{1}+\textbf{1}\otimes \hat{P}_{\mu}\label{G1}\\
		\Delta(\hat{M}_{\mu\nu})&\to \Delta_0(\hat{M}_{\mu\nu})=\hat{M}_{\mu\nu}\otimes\textbf{1}+\textbf{1}\otimes \hat{M}_{\mu\nu}\label{G2}\\
		S(\hat{P}_{\mu})&\to S_0(\hat{P}_{\mu})=-\hat{P}{\mu}\label{G3}\\
		S(\hat{M}_{\mu\nu})&\to S_0(\hat{M}_{\mu\nu})=-\hat{M}_{\mu\nu}\label{G4}
	\end{align}
	First observe that the abelian sub-algebra $[\hat{P}_{\mu},\hat{P}_{\nu}]=0$ (\ref{2}) indicates that the space $\mathcal{J}$ of abelian space-time translation generators $\hat{P}_{\mu}\in \mathcal{J} \subset \mathfrak{iso}(1,3)$ in the commutative ($a_{\mu} \to 0$) limit  $(\hat{\mathcal{M}}\to \mathcal{M})$ with $\hat{X}_{\mu} \in \hat{\mathcal{M}} \to 
	\hat{q}_{\mu} \in \mathcal{M}$ can be used to describe the standard quantum mechanical phase space, associated with undeformed Heisenberg algebra $([\hat{q}_{\mu},\hat{q}_{\nu}]=0=[\hat{P}_{\mu},\hat{P}_{\nu}];[\hat{P}_{\mu},\hat{q}_{\nu}]=-i\eta_{\mu\nu})$,
	in terms of a smash product $\mathcal{H}_0:= \mathcal{U}(\mathcal{J})\, \# \,\mathcal{U}(\mathcal{M})$, defining Heisenberg double with undeformed Heisenberg Hopf algebroid structure \footnote{As we have shown later, that $\hat{q}_{\mu}$ in our case can be related to $\hat{X}_{\nu}$ through a momentum dependent non-singular matrix \eqref{32}, whereas the momentum undergoes no deformation $P=p$.}. Note that here the 4-momenta generators which are dual to the $q$'s, can act on $\mathcal{M}$. And in this Hopf algebroid, we have two abelian Hopf algebras, given as functions of $\hat{q}^{\mu}$ and $\hat{P}_{\mu}$, which are dual to each other. The coalgebra sector of $\hat{q}_{\mu}$'s are also clearly  undeformed, therefore primitive. In particular, the coproduct of $\hat{q}_{\mu}$ is
	\begin{equation}\Delta_0(\hat{q}_{\mu})=\hat{q}_{\mu}^{(1)}\otimes \hat{q}_{\mu}^{(2)}=\hat{q}_{\mu}\otimes \textbf{1}+\textbf{1}\otimes \hat{q}_{\mu}\label{G5}
	\end{equation}
	where we have made use of the Sweedler's notation. The primitive coproduct of $\hat{P}_{\mu}$ \eqref{G1} can like-wise be expressed as $\Delta_0(\hat{P}_{\mu})=\hat{P}_{\mu}^{(1)}\otimes \hat{P}_{\mu}^{(2)}$. Using this notations, we can define the cross-multiplication rules for $\mathcal{H}_0$ as \cite{lukierski,kovacevic}
	\begin{equation}
		\hat{P}_{\mu}\hat{q}_{\nu}= \hat{q}_{\nu}^{(1)}\langle \hat{P}_{\mu}^{(1)},\hat{q}_{\nu}^{(2)}\rangle \hat{P}_{\mu}^{(2)}\label{G6}
	\end{equation}
	where $\langle . ,. \rangle$ represents canonical duality pairing and is defined as,
	\begin{equation}
		\langle \hat{P}_{\mu}, \hat{q}_{\nu}\rangle :=  \hat{P}_{\mu} \triangleright \hat{q}_{\nu} =[\hat{P}_{\mu},\hat{q}_{\nu}] \triangleright \textbf{1} =-i\eta_{\mu\nu}. \label{G7}
	\end{equation}
	And this, in a Hopf-algebraic scheme, corresponds to the binary duality map $$\mathcal{J}\otimes \mathcal{M} \to \mathbb{C}$$
	$$p \otimes q \to \langle p , q \rangle$$
	Note that, here we are just dealing with translational abelian subalgebra $\mathcal{J}$ of Poincare algebra: $\mathcal{J} \subset \mathfrak{iso}(1,3)$. Equation \eqref{G7} needs to be augmented with the following actions
	\begin{equation}
		\hat{q}_{\mu}\triangleright \textbf{1}=\hat{q}_{\mu},\quad \hat{P}_{\mu}\triangleright \textbf{1}=0\label{G8}
	\end{equation}
	Now to capture noncommutative phase space, we need to construct the deformed smash product $\mathcal{H}:=\mathcal{U}(\mathcal{\hat{J}})\,\#\,\mathcal{U}(\mathcal{\hat{M}})$, where we need to essentially replace the primitive coproduct of abelian generators \eqref{2} $\hat{P}_{\mu}$ by the deformed one: $\Delta_0(\hat{P}_{\mu})\to \Delta(\hat{P}_{\mu})$ \eqref{w1}. On the other hand, the actions of $\hat{P}_{\mu}$ and $\hat{X}_{\mu}$, along with the duality  retain their same forms given by
	\begin{equation}
		\hat{X}_{\mu} \triangleright \textbf{1} =\hat{X}_{\mu}, \quad \langle \hat{P}_{\mu}, \hat{X}_{\nu}\rangle :=  \hat{P}_{\mu} \triangleright \hat{X}_{\nu} =[\hat{P}_{\mu},\hat{X}_{\nu}] \triangleright \textbf{1} =-i\eta_{\mu\nu}
	\end{equation}
	Not only that, the cross multiplication in $\mathcal{H}$ also retains the same form \eqref{G6} 
	\begin{equation}
		\hat{P}_{\mu}\hat{X}_{\nu}= \hat{X}_{\nu}^{(1)}\langle \hat{P}_{\mu}^{(1)},\hat{X}_{\nu}^{(2)}\rangle \hat{P}_{\mu}^{(2)}\label{w2}
	\end{equation}
	except that we need to use the primitive co-product for $\hat{X}_{\mu}$ i.e. $\Delta_0(\hat{X}_{\mu})=\hat{X}_{\mu}^{(1)}\otimes \hat{X}_{\mu}^{(2)}= \hat{X}_{\mu}\otimes\textbf{   1}+\textbf{1}\otimes \hat{X}_{\mu}$ (\ref{X3}) here, rather than the deformed coproduct $\Delta(\hat{X}_{\mu})$ (\ref{X2}). To understand the reason behind this, consider the following pairing with the rule $\langle a\otimes b,c\otimes d \rangle =\langle a,c\rangle\langle b,d\rangle$
	\begin{align}
		\langle \Delta \hat{P}_{\mu}, \hat{X}_{\nu}\otimes\hat{X}_{\rho}\rangle 
		& = \langle \hat{P}_{\mu},\hat{X}_{\nu}\rangle \langle \phi,\hat{X}_{\rho}\rangle - a_{\mu }\Big\langle \phi^{-1}\Big(\hat{P}_{\lambda}-\frac{a_{\lambda}}{2} F\Big),\hat{X}_{\nu}\Big\rangle\langle \hat{P}^{\lambda},\hat{X}_{\rho}\rangle\nonumber\\
		&=-i\eta_{\mu\nu}a_{\rho}+ia_{\mu}\eta_{\rho\nu}\label{x1}
	\end{align}
	where we have used $\langle f(\hat{P}),\hat{X}_{\rho}\rangle =f(\hat{P})\triangleright \hat{X}_{\rho}=[f(\hat{P}),\hat{X}_{\rho}]\triangleright \textbf{1}$ where $f(\hat{P})$ is an arbitrary function of $\hat{P}$  and the action of $\phi(\hat{P})$ \eqref{19} on unity is given by $\phi(\hat{P})\triangleright \textbf{1}=\textbf{1}$.\\
	Now anti-symmetrizing the relation (\ref{x1}) with respect to the indices $\nu,\rho$, we can write, using \eqref{1}
	\begin{equation}
		\langle  \Delta (\hat{P}_{\mu}), \hat{X}_{\nu}\otimes\hat{X}_{\rho}- \hat{X}_{\rho}\otimes\hat{X}_{\nu}\rangle= \langle \hat{P}_{\mu}, [\hat{X}_{\nu},\hat{X}_{\rho}]\rangle=i (\eta_{\mu\rho}a_{\nu}-\eta_{\mu\nu}a_{\rho})\label{b42}
	\end{equation}
	This shows that the commutator algebra \eqref{1} is dual to the coproduct $\Delta(\hat{P}_{\mu})$. One can easily check, at this stage, that this duality will not hold if we were to make use of the undeformed i.e. primitive coproduct $\Delta_0(\hat{P}_{\mu})$ \eqref{G1} i.e. $  \langle  \Delta_0 (\hat{P}_{\mu}), \hat{X}_{\nu}\otimes\hat{X}_{\rho}- \hat{X}_{\rho}\otimes\hat{X}_{\nu}\rangle \ne \langle \hat{P}_{\mu}, [\hat{X}_{\nu},\hat{X}_{\rho}]\rangle$. On the other hand, the vanishing nature of the pairing
	\begin{equation}
		\langle \hat{P}_{\mu}\otimes \hat{P}_{\nu}- \hat{P}_{\nu}\otimes \hat{P}_{\mu}, \Delta_0(\hat{X}_{\lambda})\rangle =\langle [\hat{P}_{\mu},\hat{P}_{\nu}], \hat{X}_{\lambda}\rangle =0\label{b43}
	\end{equation}
	indicates that the vanishing commutator of $\hat{P}_{\mu}$'s \eqref{2} is dual to the primitive coproduct $\Delta_0(\hat{X}_{\mu})$. Again this duality will not hold if we were to use the deformed coproduct $\Delta(\hat{X}_{\mu})$ \eqref{X2} as this involves $\hat{P}_{\mu}$'s in its expansion \footnote{This coproduct $\Delta(\hat{X}_{\mu})$ \eqref{X2} will correspond to the Hopf algebroid $\tilde{\mathcal{H}}:= \mathcal{U}(\mathfrak{iso}(1,3)) \,\# \,\mathcal{U}(\hat{\mathcal{M}})$, where the algebra $\mathcal{A}$ is associated with enlarged basis $(\hat{X}_{\mu},\hat{P}_{\mu},\hat{M}_{\mu\nu})$ (Appendix-\ref{app2a}) \cite{lukierski}. Here,
		of course, we are not concerned with that.}.\\These important observations deserve to be emphasised once more. So to put it in other words, the commutator algebra \eqref{1} involving space-time coordinates $\hat{X}_{\mu}$ is dual \eqref{b42} to \textit{deformed} coproduct of momenta $\Delta(\hat{P}_{\mu})$ \eqref{w2}. In contrast, the vanishing commutator algebra \eqref{2} involving momenta $\hat{P}_{\mu}$ is dual \eqref{b43} only to the undeformed i.e. the primitive coproduct of $\hat{X}_{\mu}$ i.e. $\Delta_0(\hat{X}_{\mu})$ \eqref{X3}.   \\
	So finally, making use of \eqref{w2} we can write, following \cite{lukierski},
	\begin{align}
		[\hat{P}_{\mu},\hat{X}_{\nu}]&= \hat{X}_{\nu}^{(1)}\langle \hat{P}_{\mu}^{(1)},\hat{X}_{\nu}^{(2)}\rangle \hat{P}_{\mu}^{(2)}-\hat{X}_{\nu}\hat{P}_{\mu}\nonumber\\
		&= -i\eta_{\mu\nu}+ m\Big[(\Delta-\Delta_0)(\hat{P}_{\mu})(\triangleright \otimes \textbf{1}) (\hat{X}_{\nu}\otimes \textbf{1})\Big]\nonumber\\
		&= -i\eta_{\mu\nu}\phi+ia_{\mu}\hat{P}_{\nu}
	\end{align}
	reproducing the deformed Heisenberg algebra \eqref{9}.\\\\
	Now with this, the co-algebra sector gets deformed, thereby deforming the entire Hopf algebra structure. We can thus see that the action of the symmetry generators is highly deformed in the two particle sector. A natural question arises: what is the deformation, if any, in the one particle sector itself and how can it be captured? In the upcoming section we try to answer this question by developing a dynamical model, namely the Lagrangian of a massive spin-less relativistic free particle moving on $\kappa$ deformed space-time, which enjoys  the same symmetries as that of the space-time itself. Further it will be shown that the model gives rise to non-trivial momentum space geometry which, in turn, will give rise to a deformed dispersion relation encoding the deformation in the one particle sector itself.
	\section{Construction of a dynamical model invariant under deformed symmetries}
	We have already seen that there is no apparent effect of deformation in the one particle sector in $\kappa$ Minkowski space-time because of the fact that the undeformed $\mathfrak{iso}$(1,3) algebra gives an undeformed Casimir $P^2=P_{\mu}P^{\mu}$, which is expected to be good enough to label spin-less one particle states by assigning a mass $m\,(P^2=m^2)$ to it. However, a deformation in the dynamics of a relativistic free particle may arise from a deformed mass-shell condition, stemming from a curved momentum space. In this section we show how to construct a dynamical model invariant under the deformed symmetries, which yields not only the classical version of a $\kappa$ Minkowski algebra (\ref{1}), along with the phase space algebra (\ref{19}) resulting from its symplectic structure, but also incorporates the above mentioned deformed mass-shell condition affecting the dynamics. As the mass-shell condition has no impact in the symplectic structure, we take it as some generic deformation of the usual mass-shell condition to begin with and later derive its actual form from the geodesic distance in a curved momentum space, which emerges as a bi-product due to the non-commutativity in space-time.\\\\
	Here we give a systematic approach for constructing a first-order form of the Lagrangian $L_f^{\tau}$ for a relativistic free particle, in a completely algebraic approach. We demand that the Lagrangian  is invariant under the deformed symmetries (\ref{a4a},\ref{a4}) of $\kappa$ Minkowski space-time, and should produce a symplectic structure, which is nothing but the phase space algebra of $\kappa$ Minkowski space-time at the classical level. The consistency between the algebraic and dynamical approaches will become obvious on the way.  It is reassuring to note that the symmetry generators derived following N\"{o}ther's approach, i.e. from a variation of the above Lagrangian, indeed produces the same Poincare generators responsible for the respective deformed transformations
	.\\ \\
	Writing the Lagrangian of a relativistic free particle, which also respects the deformed symmetries of a $\kappa$ Minkowski space-time is nontrivial due to the Lie algebraic type of noncommutativity between the coordinates (\ref{1}) and the complicated structure of the deformation in the sector of the phase-space algebra (\ref{19}). Earlier this has been done for Snyder space-time  \cite{rabin1,rabin2} in a different approach. Here we develop a new, but straightforward, method to construct the required Lagrangian, which we now discuss step by step.\\
	The first order form of the Lagrangian for a relativistic free particle in $\kappa$ Minkowski space-time, which is a constrained system, must produce Dirac brackets consistent with the coordinate and phase space algebras corresponding to the commutators  (\ref{1}) and (\ref{a3},\ref{19}) and can be regarded as the classical counterpart of the corresponding commutators. Our starting point is therefore to demote the operators $\hat{X},\hat{P}$ to classical commuting variables $X,P$ fulfilling appropriate Dirac brackets following from the commutators (\ref{1}) and (\ref{19}) via suitable substitution.
	We are thus basically working in the limit $\hbar \to 0$. Consequently, no quantum effects or any effect of non-commutativity in the form of a fundamental length scale (like $L_p=\sqrt{\hbar G}\,(c=1)$) are expected to survive in this limit. Particularly, the non-commutative parameters $a^{\mu}$'s which were taken to be of the order of Planck length scale $L_p$ seem to vanish in this limit. One should, however, note that, the construction of the Dirac bracket from the commutator bracket involves the identification:
	\begin{equation}
		[\hat{f},\hat{g}] \longrightarrow  \{f,g\}_ {D.B}=\lim_{\hbar \to 0} \frac{1}{i\hbar}\,\,[f,g]\label{db}
	\end{equation}		
	if $\hbar$ is reinstated in the commutator brackets. We therefore have to effectively replace $a^{\mu} \to \frac{a^{\mu}}{\hbar}=:\mathfrak{a}^{\mu}$ in all the expression of (\ref{1}) and (\ref{19}) and write,
	\small{
		\begin{equation}
			\{X_{\mu},X_{\nu}\}_{D.B}=\mathfrak{a}_{\mu}X_{\nu}-\mathfrak{a}_{\nu}X_{\mu}=\theta_{\mu\nu};\,\, \{P_{\mu},X_{\nu}\}_{D.B}= -\eta_{\mu\nu}\Big[\mathfrak{a}.P+\sqrt{1+\mathfrak{a}^2P^2}\Big]+\mathfrak{a}_{\mu}P_{\nu};\,\,
			\{P_{\mu},P_{\nu}\}_{D.B}=0\label{20}\end{equation}}
	\normalsize
	Now $\mathfrak{a}^{\mu}$ will be of the order of $\sqrt{\frac{G}{\hbar}}$. Although the Planck length $\sqrt{G\hbar}$ does not survive the limit $\hbar \to 0$, the ratio $\frac{G}{\hbar}$  can be made to survive the limit by taking $G \to 0$ simultaneously, holding the ratio $\frac{G}{\hbar}$ fixed. This furnishes us with a mass scale $\sqrt{\frac{G}{\hbar}} \sim \frac{1}{\kappa}$, which, ideally should be determined experimentally, but can presumably be taken to be the inverse Planck mass $m_p$. This can be taken as a regime of quantum gravity, where there is no inherent length scale but a mass scale $m_p$. One therefore, does not anticipate to see any effect of the noncommutative nature of space-time stemming from a length scale, but can see the effect of finite mass scale through the emergence of curved momentum space \cite{smolin}, which in turn is expected to deform the dispersion relation even for a single relativistic particle.\\\\
	The method of obtaining Dirac brackets from a given constrained Lagrangian is well known in the literature \cite{hanson}.	However, here, having the expected Dirac brackets (\ref{20}) at our disposal, we trace the path backward for getting the corresponding first order Lagrangian in the following way:
	\begin{itemize}
		\item{Recall that the
			Dirac bracket (D.B) $\{.,.\}_{DB}$ between the phase space variables can be written in terms of Poission bracket $\{.,.\}$ as
			\small{
				\begin{equation}
					\{\xi_{\mu}^{(a)},\xi_{\nu}^{(b)}\}_{D.B}= \{\xi_{\mu}^{(a)},\xi_{\nu}^{(b)}\}- \{\xi_{\mu}^{(a)},\Sigma_{\alpha}^{(c)}\}(\Lambda^{-1})^{\alpha\beta}\,_{cd}\{\Sigma_{\beta}^{(d)},\xi_{\nu}^{(b)}\};\,\,\,\,\, a,b=1,2; \mu,\nu=0,1,2,3\label{DB}
			\end{equation}}
			\normalsize
			In a first order Lagrangian both $X^{\mu}$ and $P_{\mu}$ are regarded as configuration space variables of an enlarged configuration space. So it is convenient to denote them as
			$\xi^{(1)}_{\mu}=X_{\mu}$ and $\xi^{(2)}_{\mu}=P_{\mu}$. A certain first order
			Lagrangian depending on $X,P$, produces two sets of constraints relating canonical momenta with functions of the extended configuration space variables that take the  following generic structural form: 
			\begin{equation}
				\Sigma_{\mu}^1=\Pi_{\mu}^X+f_{\mu}(X,P)\approx 0;\qquad \Sigma_{\mu}^2=\Pi_{\mu}^P+g_{\mu}(X,P)\approx 0.\label{23}
			\end{equation}
			Here we have denoted $\Pi^X_{\mu} = \frac{\partial L}{\partial \dot{X}^{\mu}}$ and $\Pi^P_{\mu}=\frac{\partial L}{\partial \dot{P}^{\mu}}$ as the canonical momenta conjugate to $X_{\mu}$ and $P_{\mu}$, fulfilling the following relations,
			\begin{equation}
				\{X_{\mu},\Pi_{\nu}^X\}=\eta_{\mu\nu}= \{P_{\mu},\Pi_{\nu}^P\}\label{PB}
			\end{equation}
			and	$f_{\mu}(X,P)$, $g_{\mu}(X,P)$ are suitable functions of configuration space, which we need to determine explicitly to construct the Lagrangian of a free relativistic particle moving in $\kappa$-Minkowski space.
			Furthermore	$\Lambda^{-1}$ in (\ref{DB}) is the inverse of the constraint matrix $\Lambda$, which can be constructed from (\ref{23}) as 
			\begin{equation}
				(\Lambda_{\mu\nu})^{ab}=	\{\Sigma_{\mu}^{(a)},\Sigma_{\nu}^{(b)}\}  \label{24}
			\end{equation}
			
		}
		
		\item{As we already have the Dirac bracket relations between the phase-space variables given by (\ref{20}), we can  easily identify,  using (\ref{23}) and (\ref{PB}), the inverse of the constraint matrix $\Lambda$ from (\ref{DB}) by varying the index $a$ of $\xi^{(a)}_{\mu}$, as follows: 
			\begin{equation}
				(\Lambda^{-1})^{\mu\nu}\,_{ab} = \begin{pmatrix}
					\theta^{\mu\nu}& \eta^{\mu\nu}\phi(P)-\mathfrak{a}^{\nu}P^{\mu}\\ 
					& \\
					-\eta^{\mu\nu}\phi(P)+\mathfrak{a}^{\mu}P^{\nu}&0
				\end{pmatrix}\label{21}
		\end{equation}}
		\item{The corresponding inverse matrix $\Lambda$ can be obtained as,
			\begin{equation}
				\Lambda_{\mu\nu,ab}= \phi^{-1}(P)
				\begin{pmatrix}
					0& &-\eta_{\mu\nu}-t(P)\mathfrak{a}_{\mu}P_{\nu}\\ & & \\
					\eta_{\mu\nu}(P)+t(P)\mathfrak{a}_{\nu}P_{\mu}& &\phi^{-1}(P)\Big[\theta_{\mu\nu}+t(P)(\theta_{\mu\alpha}\mathfrak{a}^{\alpha}P_{\nu}-\theta_{\nu\alpha}\mathfrak{a}^{\alpha}P_{\mu})\Big]\end{pmatrix}\label{22}
			\end{equation}
			where $t(P)=\frac{1}{\phi(P)-\mathfrak{a}.P}$. One can indeed verify the identity  $\Lambda_{\mu\nu,ab}(\Lambda^{-1})^{\nu\lambda}\,_{bc}=\delta_{\mu}\,^{\lambda}\,\,\delta_{ac}$ holds.\\
			Our next task is to find the explicit forms of the functions $f_{\mu}(X,P)$ and $g_{\mu}(X,P)$ in the constraints (\ref{23}) , keeping in mind that the constraints should satisfy the relation (\ref{24}). Simple observations, along with some guessworks helps us to arrive at the following set of solutions for the constraints:
			\begin{equation}
				f_{\mu}(P)=0;\qquad g_{\mu}(X,P)=\phi^{-1}(P)\Big[X_{\mu}+\frac{(\mathfrak{a}.X)P_{\mu}}{\phi(P)-\mathfrak{a}.P}\Big]\label{25}
			\end{equation}
			One can check by substituting the solutions (\ref{25}) back in (\ref{23}) that the constraints (\ref{24}) are indeed satisfied and is unique upto a total time derivative (see Appendix-B).}
		\item{Once we have found the constraints, we can recover the explicit expressions of the canonical momenta as, 
			\begin{equation}
				\Pi_{\mu}^X= -f_{\mu}=0;\qquad \Pi_{\mu}^P=  -g_{\mu}=-\phi^{-1}(P)\Big[X_{\mu}+\frac{(\mathfrak{a}.X)P_{\mu}}{\phi(P)-\mathfrak{a}.P}\Big] \label{a5}
			\end{equation}
			Finally, we can write down the form of the desired Lagrangian of a relativistic free particle in first order form as	
			\begin{equation}
				L_f^{\tau}=-\phi^{-1}(P)\Big[X_{\mu}+\frac{(\mathfrak{a}.X)P_{\mu}}{\phi(P)-\mathfrak{a}.P}\Big]\dot{P}^{\mu}-e[f(P^2)-f(m^2)]\label{26}
			\end{equation}
			Here $\tau$ is the evolution parameter of the system and $e$ is a Lagrangian multiplier enforcing an anticipated deformed mass-shell condition $f(P^2)-f(m^2)=0$ where $M=\sqrt{f(m^2)}$ is to be identified with the observed renormalised mass of the particle, while $`m$' can be identified as the `bare' mass occuring in the eigen-value equation of the Casimir operator $P^2=m^2$. We assume that $P^{\mu}$ is a time-like vector and $m >0$. One can now run the Hamiltonian analysis with this form of the Lagrangian to find that the desired classical realization of the $\kappa$ Minkowski phase-space algebra (\ref{20}) is reproduced (see Appendix-B).}
	\end{itemize}
	Note that here we have taken a deformed mass-shell condition where the corresponding d'Alembertian operator is chosen to be a function of $P^2$ as $f(P^2)$ instead of just $P^2$, since we expect that it too commutes with all the generators of the $\mathfrak{iso}(1,3)$ algebra (\ref{2}). Clearly, the Hamiltonian $H$ can here be identified with
	\begin{equation}
		H=e(f(P^2)-M^2)\label{p1}
	\end{equation}
	It turns out that the exact form of the function $f$ has no bearing on the symplectic structure (\ref{20}), but has an effect on the time evolution of the system, which can now be viewed as the unfolding of gauge transformations generated by the first class constraint $f(P^2)-M^2 \approx 0$. In the next sub-section we identify the exact functional form of $f$ by relating  the d'Alembertian, with the square of the geodesic distance between a preferable origin, taken to correspond to the vacuum $P^{\mu}=0$, and a point $p$ with coordinate $P^{\mu}$ in momentum space $\mathcal{P}$, and will indeed find it to be deformed as a consequence of the curved and non-trivial nature of momentum space geometry.
	
	\subsection{Deformed mass-shell condition}
	In special theory of relativity, in absence of any length or mass scale the dispersion relation can be interpreted as the square of the straight line distance between a preferably chosen origin taken to correspond to the ground state, coordinized as $P^{\mu}=0$ to a particular point $p$ having coordinates $P^{\mu}$ in flat momentum space $\mathcal{P}_0$ (Fig-1).
	\begin{figure}[H]
		\centering
		\includegraphics[scale=0.5]{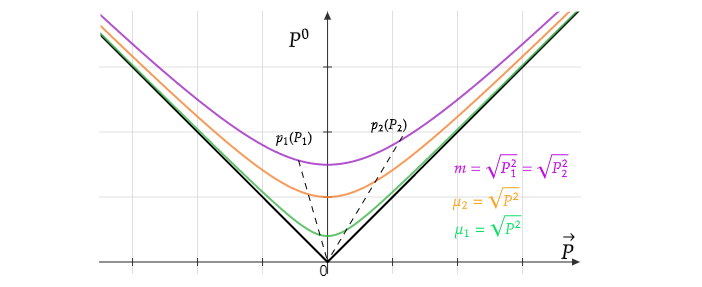}
		\caption{ One parameter family of hyperboloids, parametrized by  eigen-values of the Casimir $\sqrt{P^2}=\mu$, covering the forward light cone in the flat momentum space $\mathcal{P}_0$. A sample of 3 hyperboloids has been shown. Here the dashed lines indicate straight line paths connecting the origin to the points on the final hyperboloid which is the mass-shell $\sqrt{P^2}=m$ with momentum coordinates $P_1^{\mu},P_2^{\mu}$ in $\mathcal{P}_0$.}
	\end{figure}
	\normalsize
	The squared distance in Minkowski space, which is $C=\eta_{\alpha\beta}P^{\alpha}P^{\beta}=P^2=m^2$, generalises \cite{glikman,carmona,relancio,franchino} in curved momentum space to
	\begin{equation}
		C=[D(0,P)]^2=\Big(\int_0^P\,\,\sqrt{g_{\mu\nu}(p)dp^{\mu}dp^{\nu}}\Big)^2=\Big(\int_0^{\tau}\,\,d\tau '\, \sqrt{g_{\mu\nu}(p)\dot{p}^{\mu}\dot{p}^{\nu}}\Big)^2\,\,;\qquad \dot{p}^{\mu}=\frac{dp^{\mu}}{d\tau '}\label{g1}
	\end{equation}
	where $g_{\mu\nu}(p)$ is the momentum space metric. Here too, we take the $P^{\mu}$'s of the flat momentum space to provide a coordinate chart for the points in the forward light cone of the curved momentum space \footnote{Note that we are excluding the null directions, as we are not dealing with massless particles here.} $\mathcal{P}$. We now embark on the task of determining $g_{\mu\nu}(P)$.\\
	To begin with, note that we can re-express the Lagrangian one-form in (\ref{26}) in a more compact form as follows :
	\begin{equation}
		L_f^{\tau}d\tau= -X^{\alpha} E(P)_{\alpha}\,^{\mu} dP_{\mu}-e(f(P^2)-M^2)d\tau;\qquad E(P)_{\alpha}\,^{\mu}=\phi^{-1} \Big(\delta_{\alpha}\,^{\mu}+\frac{\mathfrak{a}_{\alpha}P^{\mu}}{\phi-\mathfrak{a}.P}\Big) \label{g2}
	\end{equation}
	Furthermore, the appearance of this $E(P)_{\alpha}\,^{\mu}$ can also be traced back to (\ref{19}), where its inverse appears as
	\begin{equation}
		\{X^{\alpha},P_{\mu}\}=(E^{-1}(P))^{\alpha}\,_{\mu};\quad (E^{-1}(P))^{\alpha}\,_{\mu}=\delta^{\alpha}\,_{\mu}\phi(P)-\mathfrak{a}_{\mu}P^{\alpha};\quad (E^{-1}(P))^{\alpha}\,_{\mu}E(P)^{\mu}\,_{\beta}=\delta^{\alpha}\,_{\beta}\label{A1} 
	\end{equation}
	in the corresponding Dirac bracket. This can be contrasted with the Lagrangian one form of a free relativistic particle in flat momentum space given by
	\begin{equation}
		\mathcal{L}_f^{\tau}d\tau=-x^{\alpha}\delta_{\alpha}\,^{\beta}dp_{\beta}-H d\tau\label{g3}
	\end{equation}
	Clearly, the possibility of curved momentum space geometry has to be encoded in the components of $E(P)$ (\ref{g2}) or $E^{-1}(P)$ (\ref{A1}), which can now be interpreted as the components of some sort of tetrad in the momentum space. It naturally suggests that a transition from the flat momentum space ($\mathcal{P}_0$) line element 
	\begin{equation}
		dS^2=\eta_{ab}dP^adP^b\label{A2}
	\end{equation}
	to the corresponding expression in curved momentum space entails suitable insertions of $E$ or $E^{-1}$ matrices, where we have made use of Latin indices deliberately to indicate the flat Lorentz indices. To proceed further, first note that $dP^a$'s in the above expression serves as an orthonormal basis of the cotangent space $T_p^*(\mathcal{P}_0)$ at an arbitrary point $p\in\mathcal{P}_0$, which is clearly holonomic in nature. The analogous orthonormal basis $e^a$'s for the cotangent space $T_p^*(\mathcal{P})$ at $p\in\mathcal{P}$, where $\mathcal{P}$ now designates the curved momentum space can only be obtained by contracting the indices of $E^{-1}$ matrix as
	\begin{equation}
		e^a=(E^{-1})^a\,_{\mu}dP^{\mu}\label{A3}
	\end{equation}
	which is generically non-holonomic (i.e. non-exact form) in nature: $de^a \ne 0$. Here we have again used Latin and Greek indices to denote orthonormal and holonomic basis elements respectively. In other words, the Latin and Greek indices now correspond to the local Lorentz and world indices respectively. 
	Note further that in the case where either $\mathfrak{a}^{\mu}$ or $P^{\mu} \to 0,\quad E_a\,^{\mu} \to \delta_a\,^{\mu}$. The line element for the curved momentum space $\mathcal{P}$ can now be easily obtained by replacing $dP^a$ in (\ref{A2}) by $e^a$ in (\ref{A3}) to get
	\begin{equation}
		dS^2=\eta_{ab}(E^{-1})^a\,_{\mu}(E^{-1})^b\,_{\nu}dP^{\mu}dP^{\nu}\label{A4}
	\end{equation}
	This helps us to just read-off the metric $\tilde{g}_{\mu\nu}(P)$ in $\mathcal{P}$ as
	\begin{equation}
		g_{\mu\nu}(P)=\eta_{ab}(E^{-1})^a\,_{\mu}(E^{-1})^b\,_{\nu}\label{A5}
	\end{equation}
	so that
	\begin{equation}
		dS^2=g_{\mu\nu}(P)dP^{\mu}dP^{\nu}\label{A6}
	\end{equation}
	holds. Now substituting $(E^{-1})^a\,_{\mu}$ from (\ref{A1}), we get
	\begin{equation}
		\tilde{g}_{\mu\nu}(\tilde{P})=\phi^2\eta_{\mu\nu}-\phi(\mathfrak{a}_{\mu}\tilde{P}_{\nu}+\mathfrak{a}_{\nu}\tilde{P}_{\mu})+\mathfrak{a}_{\mu}\mathfrak{a}_{\nu}\tilde{P}^2\label{g6}
	\end{equation}
	the inverse of which is given as
	\begin{equation}
		\tilde{g}^{\mu\nu}(\tilde{P})=\eta^{ab}E_a\,^{\mu}(\tilde{P})E_{b}\,^{\nu}(\tilde{P})=\phi^{-2}\Big[\eta^{\mu\nu}+\frac{\mathfrak{a}^{\mu}\tilde{P}^{\nu}+\mathfrak{a}^{\nu}\tilde{P}^{\mu}}{\phi-\mathfrak{a}.\tilde{P}}+\frac{\mathfrak{a}^2\tilde{P}^{\mu}\tilde{P}^{\nu}}{(\phi-\mathfrak{a}.\tilde{P})^2}\Big];\quad \tilde{g}^{\mu\nu}\tilde{g}_{\nu\rho}=\delta^{\mu}\,_{\rho}
	\end{equation}
	The significance of using overhead tilde over the metric $g_{\mu\nu}$ and $P^{\mu}$'s will become clear in the discussion below. Note that neither $\tilde{g}_{\mu\nu}$ nor $\tilde{g}^{\mu\nu}$ are covariant even under Lorentz transformation, particularly because of the presence of $\mathfrak{a}_{\mu}$'s, which are not vectors. So they are not even proper Lorentz tensors in that sense and so the momentum space $\mathcal{P}$ cannot be identified with a proper differentiable manifold, endowed with a covariantly transforming metric field $g_{\mu\nu}(P)$ under diffeomorphism. In fact, this $g_{\mu\nu}(P)$  represent a 4-parameter family of deformed metrics and $\mathcal{P}$ cannot be identified  with a group manifold or for that matter any kind of maximally symmetric space unlike \cite{glikman}, as the latter can involve only a single mass scale. However, since the geodesic distance  (\ref{g1}) in a curved manifold should be invariant under diffeomorphism, we cannot make use of non-tensorial metric $\tilde{g}_{\mu\nu}(\tilde{P})$ (\ref{g6}) given above to compute the geodesic distance. In this situation, to extract any sensible meaning about the extremal distance, we can \textit{formally} think of a covariantly transforming metric $g_{\mu\nu}(P)$ such that $\tilde{g}_{\mu\nu}(\tilde{P})$ (\ref{g6}) will be treated as a particular form of the metric $g_{\mu\nu}(P)$ in a chosen inertial frame, taken to be a fiducial frame and the corresponding quantities are  distinguished by overhead tildes. This will automatically ensure the diffeomorphism invariance of $dS^2$ \eqref{A6}. However, to do so, we need to \textit{formally} promote both $\mathfrak{a}^{\mu}$'s- a quadruplate of scalars and $P^{\mu}$'s-Lorentz four vectors to vectorially transforming objects under diffeomorphism as
	\begin{equation}
		\mathfrak{a}^{\beta} \to \frac{\partial P^{\mu}}{\partial \tilde{P}^{\beta}}\mathfrak{a}^{\beta};\qquad \tilde{P}^{\beta} \to \frac{\partial P^{\mu}}{\partial \tilde{P}^{\beta}}\tilde{P}^{\beta}.
	\end{equation} 
	Likewise we do the same for their covariantly transforming counterparts $\mathfrak{a}_{\mu},P_{\mu}$ as well, so that the metric $g_{\mu\nu}(P)$ in any frame can be obtained from $\tilde{g}_{\mu\nu}(\tilde{P})$ by requiring it to transform covariantly as
	\begin{equation}
		\tilde{g}_{\alpha\beta}(\tilde{P}) \to g_{\mu\nu}(P) = \frac{\partial \tilde{P}^{\alpha}}{\partial P^{\mu}} \frac{\partial \tilde{P}^{\beta}}{\partial P^{\nu}} \tilde{g}_{\alpha\beta}(\tilde{P})\label{A7}
	\end{equation}
	and the flat metric $\eta_{\alpha \beta}$ to transform likewise as 
	$\eta_{\alpha \beta}\rightarrow G_{\mu\nu}=\frac{\partial \tilde{P}^{\alpha}}{\partial P^{\mu}} \frac{\partial \tilde{P}^{\beta}}{\partial P^{\nu}}\eta_{\alpha\beta}$. We shall restore the respective statuses of $\mathfrak{a}^{\alpha}, P^{\alpha}$ very soon.\\\\ Note that we are using initial greek indices like $\alpha$, $\beta$ as sub/superscripts of entities refered to the fiducial frame and middle ones like $\mu$, $\nu$ etc. in any other frame. Further, at this \textit{formal} level, we can see that
	\begin{equation}
		\tilde{P}^2=\eta_{\alpha\beta}\tilde{P}^{\alpha}\tilde{P}^{\beta}=\mu^2 \to G_{\mu\nu} \frac{\partial P^{\mu}}{\partial\tilde{P}^{\alpha}} \frac{\partial P^{\nu}}{\partial\tilde{P}^{\beta}}\tilde{P}^{\alpha}\tilde{P}^{\beta}=\eta_{\alpha\beta}\tilde{P}^{\alpha}\tilde{P}^{\beta}=\mu^2
	\end{equation}
	and therefore $\mu^2$ remains invariant and can be treated as a scalar under diffeomorphism as well. Consequently,
	\begin{equation}
		\tilde{g}_{\alpha\beta}(\tilde{P})\tilde{P}^{\alpha}\tilde{P}^{\beta}=\tilde{P}^2(1+\mathfrak{a}^2\tilde{P}^2)=\mu^2(1+\mathfrak{a}^2\mu^2)\label{a9}    
	\end{equation}
	\normalsize
	is also diffeomorphism invariant. Here we have made use of the condition $\tilde{P}^2=\eta_{\alpha\beta}\tilde{P}^{\alpha}\tilde{P}^{\beta}=\mu^2$, appropriate for the hyperboloid labelled by $\sqrt{\tilde{P}^2}=\mu$ (Fig.1). This quantity (\ref{a9}) is specifically needed for our computation of geodesic distance which we shall discuss next. So it is evident that even if we were to compute this quantity (\ref{a9}) for any other choice of the fiducial frame, our results will not change.\\\\
	Now coming to the computation of the deformed dispersion relation, we first need to compute the geodesic distance between the origin, representing the ground state with $P^{\mu}=0$, and an arbitrary point $p$ with coordinate $P^{\mu}$,  in the forward light cone in the energy-momentum space $\mathcal{P}$. Clearly, the geodesic must be time-like, in the sense that the tangent vector at any point of the geodesic is necessarily time-like, but now the non-trivial nature of $\tilde{g}_{\mu\nu}$ (\ref{g6}) makes it quite difficult to compute the geodesic distance explicitly. We shall therefore adopt an alternative approach, where we make use of the differential equation \cite{thanu} satisfied by the geodesic distance $D(P):=D(0,P)$ (\ref{g1}), given by
	\begin{equation}
		\partial^{\mu}D(P)\, g_{\mu\nu}(P)\,\partial^{\nu}D(P)=1; \qquad \partial^{\mu}:=\frac{\partial}{\partial P_{\mu}}
	\end{equation}
	where we pretend as if $\mathcal{P}$ is a genuine Riemannian manifold\footnote{As indicated in \cite{smolin,glikman} that the manifold may have torsion and also the metricity condition may not hold anymore. In this sense, it may not correspond to any pseudo-Riemannian manifold; may be somewhat beyond this. But, all these issues are beyond the ambit of this paper, as this requires the study of multi-particle system and the rules for the composition of momenta of individual particles. Particularly important is the presence of any violation of commutativity and associativity of the addition of linear momenta.} .
	Using $C=D^2$, where $C$ is the d'Alembertian operator, one can express this equivalently as,
	\begin{equation}
		\partial^{\mu}C(P)\, g_{\mu\nu}(P)\,\partial^{\nu}C(P)=4C\label{g7}
	\end{equation}
	We now restrict our search for $C$ to the form $C=f(P^2)$, for reasons explained earlier. Geometrically this just means that $D(0,P_1)=D(0,P_2)$, if $P_1$ and $ P_2$ both belong to the same hyperboloid: $P_1^2=P_2^2=m^2$. In other words, we assume that this equality persists to hold even for the curved space $\mathcal{P}$ as in the case of flat space $\mathcal{P}_0$ (see Fig.1). Therefore all the points in the same mass-shell remain equidistant from the origin. This  also ensures the Poincare invariance of the entire Lagrangian (\ref{26}). Finally substituting $C=f(P^2)$ in (\ref{g7}) and using (\ref{a9})we get,
	\begin{equation}
		M=\sqrt{C}=D(0,P)=\frac{1}{2}\int_0^{P^2=m^2} \frac{d(\mu^2)}{\sqrt{\mu^2(1+\mathfrak{a}^2\mu^2)}} =\int_0^{m}\,\,\frac{d\mu}{\sqrt{ 1+\mathfrak{a}^2\mu^2}},\qquad \mathfrak{a}^2=\eta_{\mu\nu}\mathfrak{a}^{\mu}\mathfrak{a}^{\nu}\label{z5}
	\end{equation}
	At this stage, we can restore the status of $\mathfrak{a}^{\mu}$ to that of four scalar constants. But note that,  so far we have not imposed any condition on $\mathfrak{a}^2$ , but now we will consider three different cases for  $\mathfrak{a}_{\mu}$ being \textquotedblleft{null}" (i.e. $\mathfrak{a}^2=0$), \textquotedblleft{space-like}" ($\mathfrak{a}^2 <0$) and \textquotedblleft{time-like}" $(\mathfrak{a}^2 >0)$ respectively.\\\\
	\underline{\textbf{Case-1 ($\mathfrak{a}^2=0$)}}\\\\
	It follows quite trivially from (\ref{z5}) that for $\mathfrak{a}^2=0$ we find no noncommutative effect in the dispersion relation as $M=m$.\\ \\
	\underline{\textbf{Case-2 ($\mathfrak{a}^2 < 0$)}}\\\\ In this case (\ref{z5}) can be simplified to
	\begin{equation}   M=\frac{1}{\sqrt{-\mathfrak{a}^2}}\Big[\sin^{-1}(m\sqrt{-\mathfrak{a}^2})\Big]\label{z6}\end{equation}
	Taylor series expansion around the commutative limit $\mathfrak{a} \to 0$, is given by
	\begin{equation}    
		M= \frac{1}{\sqrt{-\mathfrak{a}^2}}\Big[\lambda+\frac{\lambda^2}{6}+\frac{3\lambda^4}{40}+...\Big], \qquad \textrm{for}\,\,\,\lambda=m\sqrt{-\mathfrak{a}^2} <1
	\end{equation} 
	Since $\sin^{-1}\lambda$ for $\lambda >1$ is undefined this naturally puts an upper bound on $m$ as $m < \frac{1}{\sqrt{-\mathfrak{a}^2}}$. The corresponding bound for $M$ is given by $M < \frac{\pi}{2\sqrt{-\mathfrak{a}^2}}$ (see Fig-2).\\
	\underline{\textbf{Case-3} ($\mathfrak{a}^2 > 0$)}\\
	In this case the integral (\ref{z5})  simplifies to 
	\begin{equation}
		M =\frac{1}{\mathfrak{a}}\sinh^{-1}(\mathfrak{a}m);\qquad \mathfrak{a}=\sqrt{\mathfrak{a}^2}\label{g8}
	\end{equation}
	From the above expression it is clear that, both $m$ and $M$ are not bounded as such [see Fig.2]. There is, however, a caveat. To see this we shall recall that $\sinh^{-1}(\xi)$ can be Taylor expanded for appropriate ranges of the dimensionless parameter $\xi:=\mathfrak{a}m$ as follows:
	\begin{align}
		\sinh^{-1}\xi
		= \left\{\begin{array}{lr}
			\xi-\frac{\xi^3}{6}+\frac{3\xi^5}{40}-\vartheta(\xi^7)+...\qquad\quad& \text{for} \,\,|\xi|<1 \\
			\pm\Big[ln|2\xi|+\frac{1}{4\xi^2}-\frac{3}{32\xi^4}+\vartheta(\xi^{-6})-...\Big]\qquad\quad &\text{for}\,\, \pm\xi\ge 1 \label{z2}
		\end{array} \right.
	\end{align}
	A smooth commutative limit $\xi \to 0$, requires that we only consider the case where $|\xi| <1$ in  (\ref{z2}). The other case for the range $ \xi \ge 1$ clearly does not lend itself to the desired commutative limit. It therefore implies that $\xi=1$ serves as a sort of a critical point and for $\xi <1$ one can think of constructing an appropriate power series expansion, in the spirit of perturbation theory, to obtain the corresponding noncommutative expression. The regime $\xi > 1$ (see dashed line in Fig.2) however, cannot be reached from the commutative end by any form of power series expansion. This regime ($m > \frac{1}{\mathfrak{a}}$ or equivalently $M > \frac{1}{\mathfrak{a}}\sinh^{-1}(1) \simeq \frac{0.89}{\mathfrak{a}}$ ), therefore has some \textquotedblleft{non-perturbative}" features. This is quite reminiscent of the binomial expansion of the Einstein's dispersion relation $E=\sqrt{\Vec{P}^2+m^2}$ for two regimes $|\Vec{P}| << m$ and $|\Vec{P}| >> m$ respectively. In the former case, we recover the usual expression of kinetic energy of non-relativistic particle : $E=\frac{|\vec{P}|^2}{2m}$ up to the additive rest mass energy and in the latter case we get the ultra-relativistic regime $E \sim |\vec{P}|$, where the particle is effectively massless. Here, i.e. in Einstein's special relativity, the critical point is given by $\frac{|\vec{P}|}{m} = 1$. Clearly this situation is in one to one correspondence with the above one. The non-relativistic limit here corresponds to the commutative limit in the above case while, the ultra-relativistic limit corresponds to the above mentioned  \textquotedblleft{non}-perturbative" regime. Here too one cannot obtain the dispersion relation in the ultra-relativistic domain from that of non-relativistic domain by any form of power series expansion.\\
	\begin{figure}[H]
		\centering
		\includegraphics[scale=0.5]{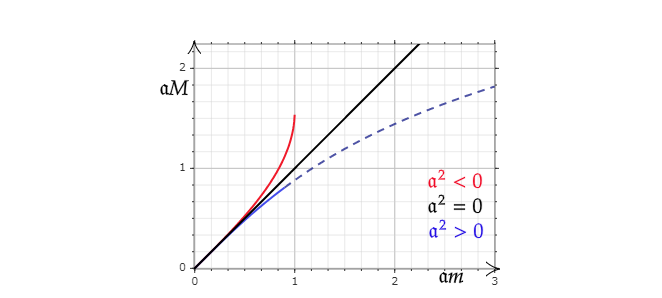}
		\caption{ Plot of $\mathfrak{a}M$ vs. $\mathfrak{a}m$ where $\mathfrak{a}:= \sqrt{\pm\mathfrak{a}^2}$ ($\pm$ for time/space like $\mathfrak{a}^{\mu}$)}
	\end{figure}
	In this context, let us recall that the fundamental reason behind the failure of any attempt to quantize gravity is that it involves a dimension full coupling constant $G$. In fact in the natural units $\frac{1}{\sqrt{G}}$ is the Planck mass/energy. Consequently, any interaction in the quantum gravity will involve power series expansion in $\frac{E}{m_P}$ and for $E > m_P$, the sum simply diverges and the theory becomes non-renormalizable. This perhaps indicates that our classical analysis of single particle kinematics will break down in the regime $\xi > 1$ (equivalently $m>\frac{1}{\mathfrak{a}}$) for \textquotedblleft{time}-like" $\mathfrak{a}^{\mu}$.\\\\
	Finally, we like to mention that the mass $M$ introduced in (\ref{z5}), can indeed be identified as renormalized mass in the spirit of Quantum field theory and can be regarded as the observed mass in contrast to the bare mass $m$. To understand this we can treat $\mathcal{P}$ as a genuine pseudo Riemannian manifold where $D^2(0,P)$ (\ref{z5}) can be recast in the following form :
	\begin{equation}
		D^2(0,P)=(\Pi^0)^2-(\Vec{\Pi})^2=\eta_{ab}\Pi^a\Pi^b=M^2
	\end{equation}
	where $\Pi^{a}$'s are the Riemann normal coordinates defined at the tangent plane $T_0(\mathcal{P})$ at the origin ${P}^{\mu}=0$ of the momentum space $\mathcal{P}$ and is related to the $P^{\mu}$'s by the well known exponential map \cite{elise}. In that case, one can have a convex neighbourhood $\mathcal{N}_0$ of $T_0(\mathcal{P})$, which maps to another convex neighbourhood $\mathcal{N}$ of $\mathcal{P}$ around the origin and one can write the following invertible relations for `space-like', `null' and `time-like'  $\mathfrak{a}$, respectively
	\begin{align}
		\Pi^a=\left\{\begin{array}{lr}\frac{P^a}{m\sqrt{-\mathfrak{a}^2}}\sin^{-1}(m\sqrt{-\mathfrak{a}^2})\qquad &\textrm{for} \quad \mathfrak{a}^2 < 0\\
			P^a \qquad\qquad\qquad\qquad\qquad\,\,\,\,\,\,\,& \textrm{for} \quad \mathfrak{a}^2 = 0\\
			\frac{P^a}{m\sqrt{\mathfrak{a}^2}}\sinh^{-1}(m\sqrt{\mathfrak{a}^2})\quad\quad\,\,\,\,\,&\textrm{for} \quad \mathfrak{a}^2 > 0 
		\end{array}\right.\label{p2}
	\end{align}
	where $n^a=\frac{P^a}{m};\,\,n^a \in T_0(\mathcal{P})$ is a unit time-like vector tangent to the geodesic at the origin: $n^an_a=\eta_{ab}n^an^b=1$. It can be parametrized as $n^a=(\cosh\psi,\sinh\psi\sin\theta\cos\phi,\\ \sinh\psi\sin\theta\sin\phi,\sinh\psi\cos\theta)$ in terms of the polar coordinates $(\psi,\theta,\phi)$ and can be used to parametrize the space-like 3D hyperboloid $n^an_a=1$ ($\mu=1$ hyperboloid in Fig.1). The momentum space $\mathcal{P}$ thus can be coordinatized in the vicinity of the origin either by a `polar coordinate' system $(\mu,\psi,\theta,\phi)$ with $\mu$ varying (Fig. 1), serving as radial coordinate or by the original $P^{\alpha}$'s themselves (which incidentally are the same as $P^a$'s occurring here) or by any other $P'^{\mu}$'s obtained by diffeomorphism transformation of $P^{\mu}$'s: $P'^{\mu}=P'^{\mu}(P^{\nu})$.\\
	Note that both the $E$ (\ref{g2}) and $E^{-1}$ (\ref{A1}) matrices reduce to identity matrices at the origin. Also in the absence of the explicit form of the geodesic equation, we cannot construct the Riemann normal coordinates at any other point on the geodesic, away from the origin. However,  $\Pi^a$'s (\ref{g6}) defined at $T_0(\mathcal{P})$ have a special status, as it is quite suggestive to identify them with the components of  renormalised and observed 4-momentum. This is further reinforced by the fact that the flat metric $\eta_{\mu\nu}$ can be retrieved from $g_{\mu\nu}(P)$ (3.22) only in the limit $P^{\mu} \to 0$ which is somewhat akin to the case where the space-time is probed by a soft photon. \\ \\
	Finally, note that all the known elementary particles are known to satisfy this kind of bound and heavier particles can be thought of being composite in nature and the total energy and momentum of such a composite system can no longer be expected to be obtained by just adding directly the momenta of the constituent particles in a generic curved momentum space \cite{arzano}. Generically, it will be less than the sum.  
	\subsection{Deformation in the form of Lorentz generators}
	To determine the deformed structures of the Lorentz generators $M_{\mu\nu}$, let us take a hint from the kinetic term $Kd\tau$ of the Lagrangian one-form (\ref{g2}) ($Kd\tau=-X^bE_b\,^{\mu}dP_{\mu}$) and introduce the so-called commutative coordinates as, 	\begin{equation}
		q^{\mu}=X^b E(P)_b\,^{\mu}=\phi^{-1}(P)\Big[X^{\mu}+\frac{(\mathfrak{a}.X)P^{\mu}}{\phi(P)-\mathfrak{a}.P}\Big]; \,\,\,\, p_{\mu}=P_{\mu}\label{32}
	\end{equation}
	so that the first term in \eqref{g2} can be expressed simply as $Kd\tau=-q^{\mu}dP_{\mu}$ , where these new phase space variables satisfy the usual i.e. undeformed phase space algebra, as one can easily check using (\ref{20})
	\begin{equation}
		\{q^{\mu},p_{\nu}\}_{D.B}=\delta^{\mu}\,_{\nu},\qquad \{q^{\mu},q^{\nu}\}_{D.B}=0\label{r9}
	\end{equation}
	The pair $(q^{\mu},p_{\nu})$ can therefore be identified as commutative variables. One can check that the inverse transformation of (\ref{32}) is simply given by,
	\begin{equation}
		X^{a}=(E^{-1}(p))^a\,_{\mu}q^{\mu}=q^{a}\phi(p)-(\mathfrak{a}.q)p^{a}\label{33}
	\end{equation}
	The occurrence of  the momentum dependent matrix $E^{-1}$ relating noncommutative and commutative coordinates, seems to be rather ubiquitous feature in various types of noncommutative space-times like the Bopp shift occurring in Moyal spaces \cite{scholtz} and Snyder space-time \cite{stern}. One must note that this mapping is given entirely in the classical level and represents a \textit{non-canonical} transformation. Consequently these mathematically defined \textquotedblleft{position-like}" coordinates $q^{\mu}$ (\ref{32}) cannot be regarded as physical position variables, unlike the original $X^{\mu}$. Nevertheless, this coordinate mapping helps us to identify the deformed Lorentz generators responsible for the deformed Lorentz transformation (\ref{a4}). To see this, note that the usual definition of Lorentz generators  of the canonical coordinates $q$ and $p$ satisfying (\ref{r9}) is given by
	\begin{equation}
		M_{\mu\nu}=q_{\mu}p_{\nu}-q_{\nu}p_{\mu}
	\end{equation}
	satisfying all the commutators given in (\ref{2}). The above transformation (\ref{32}) enables us to recast it in terms of non-commutative coordinates $X^{\mu}$, as
	\begin{equation}
		M_{\mu\nu}=\phi^{-1}(P)(X_{\mu}P_{\nu}-X_{\nu}P_{\mu})\label{a7}
	\end{equation}
	One can cross-check using (\ref{a7}) that the classical analogue of (\ref{2}) and (\ref{a3},\ref{19}) are simultaneously satisfied at the level of the classical bracket. Note that the translation generator $P$ as we have already seen remains undeformed. We shall later verify using Noether's approach (from invariance of the Lagrangian (\ref{26})), that the deformed Lorentz generator has indeed the same structure as shown in (\ref{a7}). 
	\subsection{Invariance of $L$ under deformed symmetries and N\"{o}ther generators}
	Here we verify that the Lagrangian (\ref{26}) is invariant under the deformed translation and Lorentz transformation i.e. the classical counterpart of (\ref{a4}). In other words, we show that the Lagrangian of a relativistic free particle on $\kappa$-Minkowski space also respects the same symmetry (deformed Poincare symmetry) as that of space-time itself. \\
	Under deformed translation, the infinitesimal transformations of $X,P$ in the classical setting are given  by:\\
	\begin{equation}
		\delta X^{\mu}= -\epsilon^{\alpha}\{P_{\alpha},X^{\mu}\}_{D.B}= \epsilon^{\mu}\phi(P)-(\epsilon .\mathfrak{a}) P^{\mu};\qquad \delta P_{\mu}=0\label{27}
	\end{equation}
	where $\epsilon$ is the infinitesimal translation parameter.
	Using (\ref{27}), we can check that the infinitesimal variation of the Lagrangian  (\ref{26}) is given by the following quasi-invariant form: 
	\begin{equation}
		\delta L= -\frac{d}{d\tau}(\epsilon .P);\label{28}
	\end{equation}
	so that under deformed translations the action remains invariant. We can also reproduce the translation generator from N\"{o}ther's prescription, which actually provides a connection between the algebraic way of obtaining generators and the dynamical method provided here.\\
	To see this, note that if the variation of a Lagrangian under a certain symmetry is given by a total time  derivative term such as $\delta L=\frac{dF}{d\tau}$, then the generator $G$ of the symmetry  transformation is given by
	\begin{equation}
		G= \Pi_{\mu}^\xi\delta \xi^{\mu}- F\label{a6}
	\end{equation}
	where $\xi$ denotes the configuration space variables. So in our case with the extended configuration space, the translation generator is given by \begin{equation}
		G^T= \Pi_{\mu}^{X}\delta X^{\mu}+\Pi_{\mu}^P\delta P^{\mu}+(\epsilon .P) = \epsilon. P\label{29}
	\end{equation}
	This is actually the contracted form of the generator with the corresponding parameter $\epsilon^{\mu}$ and we can thus identify the generator to be $P_{\mu}$ itself. Note that, in this article we have not introduced any intermediate translation generator \textquotedblleft{$\partial_{\mu}$}"s to represent $P_{\mu}$ as it is done in \cite{dimitrijevic,dimitrijevic2}. Unlike $\partial_{\mu}$, which has enveloping algebra valued transformation, $P_{\mu}$'s transform as four vector under Lorentz transformation. We call these $P_{\mu}$'s as our \textit{physical momentum space} in contrast to the literature referred above. \\\\
	On the other hand under deformed Lorentz transformation, the infinitesimal transformations of $X,P$ are given by
	\begin{align}
		\delta X^{\mu}&= \frac{\omega^{\alpha\beta}}{2}\{M_{\alpha\beta},X^{\mu}\}_{D.B}=\omega^{\mu}\,_{\alpha}X^{\alpha}+\omega^{\alpha\beta}\mathfrak{a}_{\alpha}M_{\beta}\,^{\mu}\nonumber\\
		\delta P_{\mu}&=\frac{\omega^{\alpha\beta}}{2}\{M_{\alpha\beta},P_{\mu}\}_{D.B}=\omega_{\mu}\,^
		{\alpha}P_{\alpha}\label{30}
	\end{align}
	With this we get a complete invariance of the Lagrangian: $\delta L =0.$
	The contracted form of the Lorentz generator can be identified in the same way as given in (\ref{a6}):
	\begin{equation}
		G^L=\phi^{-1}(P)\omega_{\mu\nu}X^{\mu}P^{\nu}=\frac{1}{2}\omega^{\mu\nu}M_{\mu\nu}\label{31}
	\end{equation}
	which exactly matches with the definition (\ref{a7}), as mentioned in the previous section.\\
	\subsection{Finite Lorentz transformation}
	The non-canonical transformations used in section-3.2, (\ref{32},\ref{33}), can formally be used to defined a finite Lorentz transformation. The finite Lorentz transformation in the commutative variables are given by
	\begin{equation}
		q^{\mu}\,\,\to\,\,q'^{\mu}=\Lambda^{\mu}\,_{\nu}q^{\nu};\qquad p_{\mu}\,\,\to\,\,p'_{\mu}=\Lambda_{\mu}\,^{\nu}p_{\nu}\label{n5}
	\end{equation}
	where $\Lambda \in SO(1,3) $. This will induce the following deformed transformation $\tilde{\Lambda}$ in the non commutative coordinates $X$'s as
	\begin{equation}
		X^{\mu}\,\to X'^{\mu}=\tilde{\Lambda}^{\mu}\,_{\nu}(\mathfrak{a},P)X^{\nu}\label{n6}
	\end{equation}
	where
	\begin{align}
		\tilde{\Lambda}^{\mu}\,_{\nu}(\mathfrak{a},P)&=\phi^{-1}(P)\Big[\phi(\Lambda P)\Lambda^{\mu}\,_{\nu}+(\Lambda P)^{\mu}(\mathfrak{a}_{\nu}-(\mathfrak{a}.\Lambda)_{\nu})\Big]\nonumber\\
		&=\Lambda^{\mu}\,_{\nu}+\phi^{-1}(P)\Big[(\mathfrak{a}.\Lambda P)\Lambda^{\mu}\,_{\nu}+(\Lambda  P)^{\mu}\mathfrak{a}_{\nu}-(\mathfrak{a}.P)\Lambda^{\mu}\,_{\nu}-(\Lambda  P)^{\mu}\mathfrak{a}_{\rho}\Lambda^{\rho}\,_{\nu}\Big]\label{p6}
	\end{align}
	while in the momentum sector the finite transformation remains undeformed.
	\begin{equation}
		P_{\mu}\,\,\to P'_{\mu}=\Lambda_{\mu}\,^{\nu}P_{\nu}
	\end{equation}
	The dependence of $\tilde{\Lambda}$ on $P_{\mu}$ and the deformation parameters $\mathfrak{a}^{\mu}$ shows that $\tilde{\Lambda}^{\mu}\,_{\nu}$'s does not close under usual multiplication unlike the undeformed $\Lambda$'s. This is indicative of the fact that the infinitesimal Lorentz transformation cannot be lifted to a finite one in this case.\\\\
	Now considering an infinitesimal Lorentz transformation $\Lambda^{\mu}\,_{\nu}=\delta^{\mu}\,_{\nu}+\omega^{\mu}\,_{\nu}$ for commutative space-time, the corresponding $\tilde{\Lambda}^{\mu}\,_{\nu}$ for the non-commutative space-time takes the following form
	\begin{equation}
		\tilde{\Lambda}^{\mu}\,_{\nu}(\mathfrak{a},P)=\delta^{\mu}\,_{\nu}+\tilde{\omega}^{\mu}\,_{\nu}(\omega,\mathfrak{a},P);\qquad \tilde{\omega}^{\mu}\,_{\nu}=\omega^{\mu}\,_{\nu}-\phi^{-1}(\omega_{\sigma\beta}\delta^{\mu}\,_{\nu}P^{\beta}-\omega_{\sigma\nu}P^{\mu})\mathfrak{a}^{\sigma},
	\end{equation}
	so that, $X^{\mu}$'s transforms as
	\begin{equation}
		X^{\mu}\,\to X'^{\mu}=X^{\mu}+\delta X^{\mu}=\tilde{\Lambda}^{\mu}\,_{\nu}(\mathfrak{a},P)X^{\nu}=(\delta^{\mu}\,_{\nu}+\tilde{\omega}^{\mu}\,_{\nu})X^{\nu}
	\end{equation}
	Interestingly this $\delta X^{\mu}$ can be simplified further to reproduce (\ref{a4}) at classical level: $\delta X^{\mu}=\frac{1}{2}\omega^{\alpha\beta}\{M_{\alpha\beta},X^{\mu}\}$. Again the infinitesimal parameter $\tilde{\omega}^{\mu}\,_{\nu}$ depends not only on the undeformed parameter $\omega$, but also on $\mathfrak{a}$ and $P$.  \\\\
	Finally note that the non-vectorial transformation properties of $X^{\mu}$'s does not give a Lorentz invariant quantity $X^{\mu}X_{\mu}=X^2$. Rather, the Lorentz invariant quantity $\mathcal{I}$ can be defined using the commutative coordinates $q^{\mu}$ as  
	\begin{equation}
		\mathcal{I}=\eta_{\mu\nu}q^{\mu}q^{\nu}=\phi^{-2}(P)\Big[X^{\mu}X_{\mu}+\frac{2(\mathfrak{a}.X)(X.P)}{\phi-\mathfrak{a}.P}+\frac{(\mathfrak{a}.X)P^2}{(\phi-\mathfrak{a}.P)^2}\Big]\label{z10}
	\end{equation}
	This is indicative of the fact that the Lorentz invariant space-time interval, in this case naturally gives a different kind of invariant interval which involves the entire phase space variables. In some sense, therefore, the whole of cotangent bundle defined on the space-time manifold plays the most fundamental role in this case. This should be contrasted with the deformed space-time interval obtained in Eq.(46) of \cite{majid3} for the case $a^0=\frac{1}{\kappa}$ and $\vec{a}=0$. And this difference stems from the deformed structure of their Poincare algebra, unlike in our case.
	\section{Conclusion and Future Outlook}
	We now summarize our findings.  We used a completely bottom-up approach for studying particle dynamics in generalized $\kappa$ Minkowski space-time. The Poincare symmetry of $\kappa$-Minkowski space-time, under which the action of a relativistic free particle also remains invariant, plays a fundamental role to get an understanding of the momentum space geometry, which in turn, helps us to derive a deformed dispersion relation. \\ While maintaining a wholly covariant framework throughout, we have first analyzed the symmetry aspects of $\kappa$ Minkowski space-time. A systematic study of Poincare symmetry has been carried out using the consistency of Jacobi identities. The result of this analysis is to obtain a set of deformed Poincare generators while the Poincare algebra itself remains unchanged. In this process, we note that the Heisenberg algebra (\ref{19}) is not the standard one and undergoes deformation. We also provide a derivation of this deformed structure using the Heisenberg-double construction in a Hopf-algebroid formulation, by following \cite{lukierski,meljanac1}. This deformation is then shown to have implications in the theory which has been unveiled through the identification of the tetrads in the momentum space $\mathcal{P}$. Following this, we give an explicit realization of a dynamical model that respects the associated symmetries by constructing an appropriate first-order Lagrangian (\ref{26}) applicable for a massive and spinless free relativistic particle in the $\kappa$ Minkowski space-time. The method we pursue here is a simple but graceful demonstration of the effectiveness of Dirac's constraint analysis or the symplectic method. The Hamiltonian is initially considered as some undetermined function of the Poincare Casimir ($P^2$), as this generalization does not interfere with the Dirac/symplectic structure of the theory. However, we find that this generalization has a lasting significance on the energy dispersion relation and in turn proves to be indeed very essential after a careful observation of the first term of our Lagrangian (\ref{26}). Here, we recognise the presence of a tetrad factor $E(P)_{\alpha}\,^{\mu}$ (\ref{g2}) in the momentum space $\mathcal{P}$ connecting the global and the local momenta variables $P^{\mu}$ and $P^a$ respectively and the closely associated emergence of non-holonomic basis $e^a$'s (\ref{A3}) in the cotangent space of the curved momentum manifold $\mathcal{P}$. This supports the existing views in the literature on the curved geometry of momentum space associated with $\kappa$-Minkowski kind of deformations of the spacetime manifold. It is quite convincing to believe that this deformation in the Heisenberg algebra (\ref{19}), obtained as a result of the consistency of the Jacobi identities, which in fact is a consequence of the stability of the $\kappa$-Minkowski algebra under the different symmetry operations, is the key factor responsible for causing deformations in the spectrum even for the free particle in Kappa spacetime. Finally, we provide a mapping between the $\kappa$-Minkowski position coordinates and the usual commutative coordinates - a reminiscent of the well-known Bopp transformations \cite{scholtz} connecting the noncommutative and the commutative coordinates in the case of the Moyal plane. We utilise this transformation to explicitly compute the deformed Lorentz generators which are subsequently verified through Noether's analysis on the Lagrangian.\\  \\
	The present analysis provides a new single particle dispersion relation that has not been reported previously. This dispersion relation may have a profound astrophysical impact as it will affect the equation of state.  In particular, repeating Chandrasekhar's analysis with this modified equation of state may influence the mass limit on white dwarfs \cite{pal}.  We also derived a relation between the mass $M$ ( namely renormalised mass) of the particle in $\kappa$- Minkowski space-time with its mass $m$ (namely the bare mass) in commutative space-time  where we have discussed three cases. For light-like deformation parameter there is no deformation in the bare mass. For space-like deformation parameter we get an upper bound in the bare mass: $m < \frac{1}{\mathfrak{a}}$ as well as for the renormalized mass $M < \frac{\pi}{2\mathfrak{a}}$. For time-like deformation parameter, $M$ becomes a monotonically increasing function of $m$. Although there is in this case no upper bound on the mass $m$ or $M$, the theory indicates the existence of a mass scale, expected to be of the order of Planck mass, beyond which growth of $M$ becomes virtually insensitive to that of $m$ and the renormalized mass sort of `saturates' in a narrow zone when $m$ is extended beyond the Planck mass scale.  This relates to the concept of relative locality \cite{smolin} which raises deep questions on the nature of space-time itself and measurements in this regime of quantum gravity. An important upshot of our analysis is that the momentum space $\mathcal{P}$ in our case cannot be identified as a group manifold, unlike \cite{glikman}. In any case, the momentum manifold $\mathcal{P}$ now involves 4 parameters $\mathfrak{a}^{\mu}$'s, which are distinct a priori. In the case where these parameters/scales differ significantly, the manifold $\mathcal{P}$ cannot correspond to a maximally symmetric space.  Even when, $\mathfrak{a}^{\mu}$ is purely time-like in the sense that $\mathfrak{a}^0 \ne 0, \vec{\mathfrak{a}} =0$, we don't recover the group manifold AN(3) of \cite{glikman}. Presumably, this difference stems from the Lorentz covariant transformation properties of the $P^{\mu}$'s in our case, in contrast to the ones of \cite{glikman}.  It thus becomes difficult to compose individual momenta of a multi-particle system to obtain the total momentum \cite{arzano}. It will therefore be quite interesting to see how one can construct multi-particle actions in the presence of interactions (taken to be simple collisions) and study the corresponding Hopf algebra symmetry.\\\\
	On the other hand, as a result of dynamical analysis for deformed Poincare symmetry, it is possible to see that the action of the finite Lorentz transformation on the $\kappa$ Minkowski space time provides an invariant quantity (\ref{z10}) which is a function of phase space variables. This may be  interpreted as a generalization of the conventional space time interval. This clearly  indicates that the geometry of the whole cotangent bundle will be the most fundamental object in this scenario, because introduction of a mass-scale unifies the coordinate space and momentum space, just as in special relativity, the introduction of a universal speed unifies space and time \cite{smolin}. This represents a paradigm shift from usual geometrical standpoint and may have a connection with  Finsler geometry \cite{finsler}.\\\\
	Finally, our proposed geometric implication of deformed space-time symmetries provides a possible resolution to some problems. For example, it is important to investigate the QG correction to the world line path-integral formulation of effective quantum field theory \cite{strassler}, which is based on the classical action for a relativistic charged particle in the presence of background gauge potential.\\
	It would also be interesting to take up the case of a relativistic spinning particle in $\kappa$-Minkowski background where the geometry of the space will no longer be Riemannian due to the induction of torsion from spinning particles.  These are some exciting questions that we leave for further studies.
	\section*{Acknowledgement}
	PN and SKP, would like to extend their gratitude to S.N. Bose National Centre for Basic Sciences, Kolkata for visiting fellowships during the initial stages of the work. One of the authors, PN,  also would like to express his gratitude to Stellenbosch University for providing postdoctoral funds during the last stage of the work. AC and BC thank Prof. A.P. Balachandran and Prof. Kumar S. Gupta for their critical comments and useful discussion. The authors would also like to thank the referee for his/her useful and constructive comments.
	\begin{appendices}
		\section{Calculation of the coproduct formulae}\label{app2a}
		Here we briefly sketch the method of obtaining the co-product for the generators $\hat{M}_{\mu\nu}, \hat{P}_{\mu}$. The action of $\hat{P}_{\mu}$ on a product of operator $\hat{X}$ valued functions $\hat{f}(\hat{X}).\hat{g}(\hat{X})$ can be shown to be given by
		\begin{align}
			\hat{P}_{\mu}\triangleright (\hat{f}(\hat{X})\hat{g}(\hat{X}))& = ([\hat{P}_{\mu},\hat{f}(\hat{X})]) \triangleright \hat{g}(\hat{X}) + (\hat{f}(\hat{X})\hat{P}_{\mu})\triangleright \hat{g}(\hat{X})\nonumber\\
			&= ([\hat{P}_{\mu}, \hat{f}(\hat{X})]) \triangleright \hat{g}(\hat{X}) + m((1 \otimes \hat{P}_{\mu}) \triangleright ( \hat{f}(\hat{X})\otimes  \hat{g}(\hat{X}))
		\end{align}
		where the sign $\triangleright$ denotes action of $\hat{P}_{\mu}$, given as
		\begin{equation}
			\hat{P}_{\mu}\triangleright \hat{f}(\hat{X}) =[\hat{P}_{\mu},\hat{f}(\hat{X})]\triangleright \textbf{1} \qquad \textrm{and} \quad \hat{P}_{\mu}\triangleright \textbf{1}=0\label{X1}
		\end{equation}
		and $m$ is the multiplication map such that $m( f\otimes g)=f.g$ and one can immediately identify that
		\begin{equation}
			m\Big(\Delta \hat{P}_{\mu} \triangleright (\hat{f}(\hat{X}) \otimes \hat{g}(\hat{X})\Big) = [\hat{P}_{\mu},\hat{f}(\hat{X})] \triangleright \hat{g}(\hat{X}) + m((1\otimes \hat{P}_{\mu})\triangleright (\hat{f}(\hat{X})\otimes \hat{g}(\hat{X}))
		\end{equation}
		So the coproduct of $\hat{P}_{\mu}$ can be obtained from the commutator relation $[\hat{P}_{\mu}, \hat{f}(\hat{X})]$. Now first taking $\hat{f}(\hat{X})=\hat{X}_{\nu}$ as a simple example and using the relation (\ref{19}) we obtain
		\begin{align}
			[\hat{P}_{\mu},\hat{X}_{\nu}] \triangleright \hat{g}(\hat{X})&=[-i\eta_{\mu\nu}\phi+ia_{\mu}\hat{P}_{\nu}] \triangleright \hat{g}(\hat{X})\nonumber\\
			& =[-i\eta_{\mu\nu}\phi+ia_{\mu}\phi^{-1}(\phi\eta_{\alpha\nu})\hat{P}^{\alpha}] \triangleright \hat{g}(\hat{X})\nonumber\\
			&= (\hat{P}_{\mu}\triangleright \hat{X}_{\nu}) \phi  \triangleright \hat{g}(\hat{X}) -a_{\mu}\Big[\phi^{-1}(\hat{P}_{\alpha}-\frac{a_{\alpha}}{2}F(\hat{P})), \hat{X}_{\nu}\Big]\triangleright \textbf{1}( \hat{P}^{\alpha} \triangleright \hat{g}(\hat{X}))\nonumber\\
			&=(\hat{P}_{\mu}\triangleright \hat{X}_{\nu})\phi  \triangleright \hat{g}(\hat{X})-a_{\mu}\Big[\phi^{-1}(\hat{P}_{\alpha}-\frac{a_{\alpha}}{2}F(\hat{P})) \triangleright \hat{X}_{\nu}\Big] \hat{P}^{\alpha} \triangleright \hat{g}(\hat{X})
		\end{align}
		where we have used the following relations,
		\begin{equation}
			[\hat{P}_{\mu},\hat{X}_{\nu}] \triangleright \textbf{1}=-i\eta_{\mu\nu},\qquad \Big[\hat{P}_{\alpha}-\frac{a_{\alpha}}{2}F(\hat{P}), \hat{X}_{\nu}\Big]=-i\eta_{\alpha\nu}\phi(\hat{P})
		\end{equation}
		where $F(\hat{P})=\frac{2}{a^2}(1-\sqrt{1+a^2\hat{P}^2})$.
		We can now generalize the above relations to an arbitrary function $\hat{f}(\hat{X})$ using the method of induction on the monomials of $\hat{X}_{\mu} $ \cite{juric, kovacevic} and write
		\begin{equation}
			[\hat{P}_{\mu},\hat{f}(\hat{X})]=(\hat{P}_{\mu}\hat{f})\phi-a_{\mu}\Big(\hat{P}_{\nu}\phi^{-1}\hat{f}\Big)\hat{P}^{\nu}+\frac{a_{\mu}}{2}\Big(F(\hat{P})\phi^{-1}\hat{f}\Big)(a.\hat{P})
		\end{equation}
		So one can finally write the coproduct of $\hat{P}_{\mu}$ as
		\begin{equation*}
			\Delta (\hat{P}_{\mu})= \hat{P}_{\mu}\otimes\phi+\textbf{1}\otimes \hat{P}_{\mu}-a_{\mu}(\hat{P}_{\nu}\phi^{-1})\otimes \hat{P}^{\nu}+\frac{a_{\mu}}{2}(F(\hat{P})\phi^{-1})\otimes (a.\hat{P})
		\end{equation*}
		Now one can also derive the coproduct of $\hat{M}_{\mu\nu}$ simply by using its realization given in (\ref{a7}) as
		\begin{align*}
			&   \Delta \hat{M}_{\mu\nu} \triangleright (\hat{f}\otimes \hat{g})=\Delta\Big((X_{\mu}\hat{P}_{\nu}-X_{\nu}\hat{P}_{\mu})\phi^{-1}(\hat{P})\Big)\triangleright (\hat{f}\otimes \hat{g})\\
			& = (\hat{M}_{\mu\nu}\hat{f})\otimes\hat{g}+\hat{f}\otimes (\hat{M}_{\mu\nu}\hat{g})+ \Big[a_{\mu}\Big(\hat{P}^{\lambda}-\frac{a^{\lambda}}{2}F(\hat{P})\Big)\phi^{-1}\Big]\hat{f}\otimes (\hat{M}_{\lambda\nu}\hat{g})-\\
			&\Big[a_{\nu}\Big(\hat{P}^{\lambda}-\frac{a^{\lambda}}{2}F(\hat{P})\Big)\phi^{-1}\Big]\hat{f}\otimes (\hat{M}_{\lambda\mu}\hat{g})   
		\end{align*}
		which gives us the coproduct formula for $\hat{M}_{\mu\nu}$ stated in (\ref{n2}).\\\\
		\textbf{Co-associativity:} One of the most important property of a Hopf algebra $H$ is that it should be co-associative i.e. the coproduct $\Delta$ should satisfy
		\begin{equation}
			(\Delta \otimes \textbf{1}) \Delta = (\textbf{1} \otimes \Delta) \Delta : H \to H\otimes H\otimes H
		\end{equation}
		For example we can check via a lengthy but straight forward calculation, the following identity:
		\begin{align}
			(\Delta \otimes \textbf{1})\Delta \hat{P}_{\mu} &= \hat{P}_{\mu} \otimes \phi \otimes \phi +\textbf{1}\otimes \hat{P}_{\mu} \otimes\phi -a_{\mu} \hat{P}_{\nu}\phi^{-1}\otimes P^{\nu} \otimes \phi +\frac{a_{\mu}}{2} F(\hat{P})\phi^{-1}\otimes a.\hat{P}\otimes \phi \nonumber\\ &+\textbf{1}\otimes \textbf{1} \otimes \hat{P}_{\mu}
			-a_{\mu}\hat{P}_{\nu}\phi^{-1}\otimes \textbf{1}\otimes \hat{P}^{\nu}-a_{\mu}\phi^{-1}\otimes \hat{P}_{\nu}\phi^{-1} \otimes \hat{P}^{\nu}\nonumber\\
			+\frac{a_{\mu}}{a^2}\Big[\phi^{-1}&\otimes\phi^{-1}\otimes a.\hat{P}-\textbf{1}\otimes\textbf{1}\otimes a.\hat{P}
			+ a.\hat{P} \phi^{-1} \otimes\textbf{1}\otimes a.\hat{P} +\phi^{-1}\otimes a.\hat{P}\phi^{-1}\otimes a.\hat{P}\Big]\nonumber\\
			&= (\textbf{1}\otimes \Delta) \Delta \hat{P}_{\mu}
		\end{align}
		To prove the above identity we have used $F(\hat{P})=\frac{2}{a^2}(1-\sqrt{1+a^2\hat{P}^2})=\frac{2}{a^2}(1-\phi+a.\hat{P})$ and the homomorphism of the coproduct i.e. $\Delta (a.b)=\Delta(a).\Delta(b),\,\,\forall a,b \in H$. We have also used the coproduct $\Delta (\phi^{-1})=\phi^{-1}\otimes \phi^{-1}$, so $\phi^{-1}$ can be regarded as a group-like element of the Hopf algebra as $\phi$ itself i.e. $\Delta(\phi)=\phi\otimes\phi$. The coproduct of $\hat{M}_{\mu\nu}$ can also be verified to be co-associative: $(\Delta\otimes\textbf{1})\Delta \hat{M}_{\mu\nu}=(\textbf{1}\otimes\Delta)\Delta \hat{M}_{\mu\nu}$ . \\
		It is important to notice that the set of commutator relations in (\ref{1},\ref{2}) and (\ref{a3},\ref{19})  closes on the universal enveloping algebra generated by the enlarged set $\{\hat{M}_{\mu\nu}, \hat{P}_{\mu}, X_{\mu}\}$. We have already discussed the coalgebra sector corresponding to the Poincare generators in section-2.1. It should now be augmented by the co-algebra structure of $\hat{X}_{\mu}$ itself. We, however, contend ourselves just by quoting the expression of coproduct of $\hat{X}_{\mu}$, given in \cite{kovacevic} as,
		\begin{equation}
			\Delta \hat{X}_{\mu}=\phi\otimes\hat{X}_{\mu}+\hat{X}_{\mu}\otimes\textbf{1}-a_{\mu}\Big[\hat{P}_{\alpha}-\frac{a_{\alpha}}{2}F\Big]\otimes \hat{X}^{\alpha}.\label{X2}
		\end{equation}
		This is because this expression is not used in the construction of Heisenberg double (see sec-2.1) on the way to provide an alternative `derivation' of deformed Heisenberg algebra (\ref{19}), for reasons explained there. In fact we have to make use of the undeformed i.e. the primitive form of coproduct :
		\begin{equation}
			\Delta_0(\hat{X}_{\mu})=\hat{X}_{\mu}\otimes\textbf{   1}+\textbf{1}\otimes \hat{X}_{\mu},\label{X3}
		\end{equation}
		as only translational generators $\hat{P}_{\mu} \in \hat{\mathcal{J}}$ will be used in the Heisenberg double construction, $\mathcal{H}=\mathcal{U}(\hat{\mathcal{J}})\,\#\, \mathcal{U}(\hat{\mathcal{M}})$, where the Lorentz generators $\hat{M}_{\mu\nu}$ will be excluded.
		\section{Symplectic analysis}\label{app2b}
		Let us first write a general first-order Lagrangian as,
		\begin{equation}
			L_f=a_i(\xi)\dot{\xi}^i-H_c\label{r7}
		\end{equation}
		where $\xi_i$ are the phase space variables, $i=1,...,2N$ for a $N$ dimensional coordinate space. Note that $a_i(\xi)$ can be considered as a sort of vector potential (connection) for an abelian gauge theory, since modification of $a_i$ by a total derivative as
		\begin{equation}
			a_i(\xi) \to  a_i(\xi) +\frac{\partial\theta}{\partial\xi^i}\label{r8}
		\end{equation}
		does not affect dynamics, since the Lagrangian changes by a total time-derivative term.  However, the structure of constraints will vary under the transformation (\ref{r8})  as can be seen directly from (\ref{r7}). \\
		Now for a generic $a_i(\xi)$, one can find out the Euler-Lagrange equation from (\ref{r7}) as 
		\begin{equation}
			f_{ij}(\xi)\dot{\xi}^j=\frac{\partial H_c}{\partial\xi^i}
		\end{equation}   
		where $f_{ij}=\frac{\partial a_j(\xi)}{\partial\xi_i}-\frac{\partial a_i(\xi)}{\partial\xi_j}$ acts as gauge invariant two form (curvature) constructed out of gauge variant connection $a_i(\xi)$ and is called symplectic two form. It can be shown that $f_{ij}(\xi)$ is basically the constraint matrix, which remains invariant under the gauge transformation of $a_i(\xi)$.\\\\
		Now we carry out the symplectic analysis \textit{a la'} Fadeev Jackiw (FJ) for the system Lagrangian and show that it indeed produces the same symplectic brackets as that of (\ref{20}), as expected. Symplectic approach is an alternative and sometimes quicker method (than Dirac's analysis) specially for first order system to obtain the phase space brackets from equations of motion. We can calculate the Euler Lagrange equation of motion from the system Lagrangian (\ref{26}) as
		\begin{align}
			\phi^{-1}\dot{P}_{\mu}+\frac{\phi^{-1}\mathfrak{a}_{\mu}}{\phi-\mathfrak{a}.P}(P.\dot{P})&=0\\
			\phi^{-1}\dot{X}_{\mu}+\frac{\phi^{-1}}{\phi-\mathfrak{a}.P}(\mathfrak{a}.\dot{X})P_{\mu}+\phi^{-2}[\mathfrak{a}_{\mu}(X.\dot{P})-X_{\mu}(\mathfrak{a}.\dot{P})&] \nonumber\\
			+\frac{\phi^{-2}}{\phi-\mathfrak{a}.P}[\mathfrak{a}_{\mu}(\mathfrak{a}.X)(P.\dot{P})-\mathfrak{a}^2X_{\mu}(P.\dot{P})-(\mathfrak{a}.\dot{P})(\mathfrak{a}.X)P_{\mu}+\mathfrak{a}^2(X.\dot{P})P_{\mu}]&=e\,\,\frac{\partial f(P^2)}{\partial P^{\mu}}\label{r5}
		\end{align}
		From FJ analysis, it is known that these equations of motion can be recast in the following form 
		\begin{equation}
\Lambda_{\mu\nu,ab}\dot{\xi}_{b,\nu}=\frac{\partial H_c}{\partial\xi^{a,\mu}}\label{r6}
		\end{equation}
		where $\Lambda_{ab,\mu\nu}$ is basically the constraint matrix. So we can read off the components of constraint matrix from (\ref{r6}) using (\ref{r5}), which exactly matches with (\ref{22}). Now the symplectic bracket between two variables $f$ and $g$ is given by 
		\begin{equation}
			\{f,g\}_{SB}=(\Lambda^{-1})^{\mu\nu}\,_{ab}\,\,\partial_{\mu,a}f\,\,\partial_{\nu,b}\,g
		\end{equation}
		where $\partial_{\mu,a}=\frac{\partial}{\partial\xi^{\mu}_a}$ and $\Lambda^{-1}$ is the inverse of the constraint matrix given in (\ref{21}). So the symplectic brackets between phase space variables are given by
		\begin{equation}
			\{X_{\mu},X_{\nu}\}_{S.B}=(\mathfrak{a}_{\mu}X_{\nu}-\mathfrak{a}_{\nu}X_{\mu})=\theta_{\mu\nu};\quad \{P_{\mu},X_{\nu}\}_{S.B}= -\eta_{\mu\nu}\phi(P)+\mathfrak{a}_{\mu}P_{\nu};\quad
			\{P_{\mu},P_{\nu}\}_{S.B}=0
		\end{equation} 
		It can be seen that, the symplectic brackets are identical with the Dirac brackets and produce the classical version of the $\kappa$ Minkowski algebra (\ref{1}).
	\end{appendices}
	\bibliographystyle{JHEP.bst}
	\bibliography{main.bib}

\providecommand{\href}[2]{#2}\begingroup\raggedright\begin{thebibliography}{10}

\bibitem{abbott}
B.P.~Abbott, \textit{et. al.} (LIGO Scientific~Collaboration and
  V.~Collaboration), \emph{Observation of gravitational waves from a binary
  black hole merger},
  \href{https://doi.org/10.1103/PhysRevLett.116.061102}{\emph{Phys. Rev. Lett.}
  {\bfseries 116} (2016) 061102}.

\bibitem{penrose}
R.~Penrose, \emph{Gravitational collapse and space-time singularities},
  \href{https://doi.org/10.1103/PhysRevLett.14.57}{\emph{Phys. Rev. Lett.}
  {\bfseries 14} (1965) 57}.

\bibitem{elise}
S.W.~Hawking and G.F.R.~Ellis, \emph{The Large Scale Structure of Space-Time},
  Cambridge Monographs on Mathematical Physics, Cambridge University Press
  (1973),
  \href{https://doi.org/10.1017/CBO9780511524646}{10.1017/CBO9780511524646}.

\bibitem{doplicher}
S.~Doplicher, K.~Fredenhagen and J.E.~Roberts, \emph{Spacetime quantization
  induced by classical gravity}, {\emph{Phys. Lett. B} {\bfseries 331} (1994)
  39}.

\bibitem{seiberg}
N.~Seiberg and E.~Witten, \emph{{String theory and noncommutative geometry}},
  \href{https://doi.org/10.1088/1126-6708/1999/09/032}{\emph{JHEP} {\bfseries
  09} (1999) 032} [\href{https://arxiv.org/abs/hep-th/9908142}{{\ttfamily
  hep-th/9908142}}].

\bibitem{vedral}
C.~Marletto and V.~Vedral, \emph{Why we need to quantise everything, including
  gravity}, {\emph{npj Quantum Information} {\bfseries 3} (2017) }.

\bibitem{hawking}
S.~Hawking, \emph{Particle creation by black holes}, {\emph{Comm.Math. Phys.}
  {\bfseries 3} (1975) 199}.

\bibitem{bose}
S.~Bose, A.~Mazumdar, M.~Schut and M.~Toro\v{s}, \emph{{Mechanism for the
  quantum natured gravitons to entangle masses}},
  \href{https://doi.org/10.1103/PhysRevD.105.106028}{\emph{Phys. Rev. D}
  {\bfseries 105} (2022) 106028}
  [\href{https://arxiv.org/abs/2201.03583}{{\ttfamily 2201.03583}}].

\bibitem{marletto}
C.~Marletto and V.~Vedral, \emph{Witness gravity’s quantum side in the lab},
  {\emph{Nature} {\bfseries 547} (2017) 156}.

\bibitem{marletto2}
C.~Marletto and V.~Vedral, \emph{Gravitationally induced entanglement between
  two massive particles is sufficient evidence of quantum effects in gravity},
  \href{https://doi.org/10.1103/PhysRevLett.119.240402}{\emph{Phys. Rev. Lett.}
  {\bfseries 119} (2017) 240402}.

\bibitem{luk2}
J.~Lukierski, H.~Ruegg, A.~Nowicki and V.N.~Tolstoy, \emph{q-deformation of
  poincaré algebra},
  \href{https://doi.org/https://doi.org/10.1016/0370-2693(91)90358-W}{\emph{Phys.
  Lett. B} {\bfseries 264} (1991) 331}.

\bibitem{luk1}
J.~Lukierski, A.~Nowicki and H.~Ruegg, \emph{New quantum poincaré algebra and
  $\kappa$-deformed field theory},
  \href{https://doi.org/https://doi.org/10.1016/0370-2693(92)90894-A}{\emph{Phys.
  Lett. B} {\bfseries 293} (1992) 344}.

\bibitem{majid3}
S.~Majid and H.~Ruegg, \emph{Bicrossproduct structure of $\kappa$-poincare
  group and non-commutative geometry},
  \href{https://doi.org/https://doi.org/10.1016/0370-2693(94)90699-8}{\emph{Phys.
  Lett. B} {\bfseries 334} (1994) 348}.

\bibitem{luk4}
J.~Lukierski, H.~Ruegg and W.~Zakrzewski, \emph{Classical and quantum mechanics
  of free $\kappa$-relativistic systems},
  \href{https://doi.org/https://doi.org/10.1006/aphy.1995.1092}{\emph{Ann.
  Phys.} {\bfseries 243} (1995) 90}.

\bibitem{lukierski}
J.~Lukierski, D.~Meljanac, S.~Meljanac, D.~Pikutic and M.~Woronowicz,
  \emph{{Lie-deformed quantum Minkowski spaces from twists: Hopf-algebraic
  versus Hopf-algebroid approach}},
  \href{https://doi.org/10.1016/j.physletb.2017.12.007}{\emph{Phys. Lett. B}
  {\bfseries 777} (2018) 1} [\href{https://arxiv.org/abs/1710.09772}{{\ttfamily
  1710.09772}}].

\bibitem{camelia/dsr}
G.~Amelino-Camelia, \emph{{Testable scenario for relativity with minimum
  length}}, \href{https://doi.org/10.1016/S0370-2693(01)00506-8}{\emph{Phys.
  Lett. B} {\bfseries 510} (2001) 255}
  [\href{https://arxiv.org/abs/hep-th/0012238}{{\ttfamily hep-th/0012238}}].

\bibitem{camelia/dsr2}
G.~Amelino-Camelia, \emph{{Relativity in space-times with short distance
  structure governed by an observer independent (Planckian) length scale}},
  \href{https://doi.org/10.1142/S0218271802001330}{\emph{Int. J. Mod. Phys. D}
  {\bfseries 11} (2002) 35}
  [\href{https://arxiv.org/abs/gr-qc/0012051}{{\ttfamily gr-qc/0012051}}].

\bibitem{glikmannew}
J.~Kowalski-Glikman, \emph{Introduction to doubly special relativity},  in
  \emph{Planck Scale Effects in Astrophysics and Cosmology}, (Berlin,
  Heidelberg), pp.~131--159, Springer Berlin Heidelberg (2005),
  \href{https://doi.org/10.1007/11377306\_5}{https://doi.org/10.1007/11377306\_5}.

\bibitem{smolin}
G.~Amelino-Camelia, L.~Freidel, J.~Kowalski-Glikman and L.~Smolin, \emph{{The
  principle of relative locality}},
  \href{https://doi.org/10.1103/PhysRevD.84.084010}{\emph{Phys. Rev. D}
  {\bfseries 84} (2011) 084010}
  [\href{https://arxiv.org/abs/1101.0931}{{\ttfamily 1101.0931}}].

\bibitem{freidel1}
G.~Amelino-Camelia, L.~Freidel, J.~Kowalski-Glikman and L.~Smolin,
  \emph{{Relative locality: A deepening of the relativity principle}},
  \href{https://doi.org/10.1142/S0218271811020743}{\emph{Gen. Rel. Grav.}
  {\bfseries 43} (2011) 2547}
  [\href{https://arxiv.org/abs/1106.0313}{{\ttfamily 1106.0313}}].

\bibitem{born}
M.~Born, \emph{A suggestion for unifying quantum theory and relativity, proc.
  r. soc. london a 165, 291 (1938)}, .

\bibitem{camelia-shahn}
G.~Amelino-Camelia and S.~Majid, \emph{{Waves on noncommutative space-time and
  gamma-ray bursts}},
  \href{https://doi.org/10.1142/S0217751X00002777}{\emph{Int. J. Mod. Phys. A}
  {\bfseries 15} (2000) 4301}
  [\href{https://arxiv.org/abs/hep-th/9907110}{{\ttfamily hep-th/9907110}}].

\bibitem{majid}
S.~Majid, \emph{Meaning of noncommutative geometry and the planck-scale quantum
  group},  in \emph{Towards Quantum Gravity}, (Berlin, Heidelberg),
  pp.~227--276, Springer Berlin Heidelberg, 2000.

\bibitem{witten}
E.~Witten, \emph{{(2+1)-Dimensional Gravity as an Exactly Soluble System}},
  {\emph{Nucl. Phys. B 311 (1988) 46} {\bfseries 311} (1988) 46}.

\bibitem{simone}
L.~Freidel and S.~Speziale, \emph{{On the relations between gravity and BF
  theories}}, \href{https://doi.org/10.3842/SIGMA.2012.032}{\emph{SIGMA}
  {\bfseries 8} (2012) 032} [\href{https://arxiv.org/abs/1201.4247}{{\ttfamily
  1201.4247}}].

\bibitem{glikman1}
J.~Kowalski-Glikman, \emph{De sitter space as an arena for doubly special
  relativity},
  \href{https://doi.org/https://doi.org/10.1016/S0370-2693(02)02762-4}{\emph{Phys.
  Lett. B} {\bfseries 547} (2002) 291}.

\bibitem{glikman2}
J.~Kowalski-Glikman and S.~Nowak, \emph{{Doubly special relativity and de
  Sitter space}},
  \href{https://doi.org/10.1088/0264-9381/20/22/006}{\emph{Class. Quant. Grav.}
  {\bfseries 20} (2003) 4799}
  [\href{https://arxiv.org/abs/hep-th/0304101}{{\ttfamily hep-th/0304101}}].

\bibitem{koch}
F.~Koch and E.~Tsouchnika, \emph{{Construction of $\theta$-Poincare algebras
  and their invariants on $M_{\theta}$}},
  \href{https://doi.org/10.1016/j.nuclphysb.2005.04.019}{\emph{Nucl. Phys. B}
  {\bfseries 717} (2005) 387}
  [\href{https://arxiv.org/abs/hep-th/0409012}{{\ttfamily hep-th/0409012}}].

\bibitem{jerzy}
J.~Kowalski-Glikman and A.~Starodubtsev, \emph{Effective particle kinematics
  from quantum gravity},
  \href{https://doi.org/10.1103/PhysRevD.78.084039}{\emph{Phys. Rev. D}
  {\bfseries 78} (2008) 084039}.

\bibitem{juric}
S.~Kresic-Juric, S.~Meljanac and M.~Stojic, \emph{{Covariant realizations of
  kappa-deformed space}},
  \href{https://doi.org/10.1140/epjc/s10052-007-0285-8}{\emph{Eur. Phys. J. C}
  {\bfseries 51} (2007) 229}
  [\href{https://arxiv.org/abs/hep-th/0702215}{{\ttfamily hep-th/0702215}}].

\bibitem{twist}
J.~Lukierski, D.~Meljanac, S.~Meljanac, D.~Pikutic and M.~Woronowicz,
  \emph{{Lie-deformed quantum Minkowski spaces from twists: Hopf-algebraic
  versus Hopf-algebroid approach}},
  \href{https://doi.org/10.1016/j.physletb.2017.12.007}{\emph{Phys. Lett. B}
  {\bfseries 777} (2018) 1} [\href{https://arxiv.org/abs/1710.09772}{{\ttfamily
  1710.09772}}].

\bibitem{meljanac2}
S.~Meljanac, A.~Samsarov, M.~Stojic and K.S.~Gupta, \emph{{Kappa-Minkowski
  space-time and the star product realizations}},
  \href{https://doi.org/10.1140/epjc/s10052-007-0450-0}{\emph{Eur. Phys. J. C}
  {\bfseries 53} (2008) 295} [\href{https://arxiv.org/abs/0705.2471}{{\ttfamily
  0705.2471}}].

\bibitem{woronowicz}
J.~Lukierski, D.~Meljanac, S.~Meljanac, D.~Pikutic and M.~Woronowicz,
  \emph{{Lie-deformed quantum Minkowski spaces from twists: Hopf-algebraic
  versus Hopf-algebroid approach}},
  \href{https://doi.org/10.1016/j.physletb.2017.12.007}{\emph{Phys. Lett. B}
  {\bfseries 777} (2018) 1} [\href{https://arxiv.org/abs/1710.09772}{{\ttfamily
  1710.09772}}].

\bibitem{dimitrijevic}
M.~Dimitrijevic, L.~Jonke, L.~Moller, E.~Tsouchnika, J.~Wess and
  M.~Wohlgenannt, \emph{{Field theory on kappa-spacetime}},
  \href{https://doi.org/10.1007/s10582-004-9785-z}{\emph{Czech. J. Phys.}
  {\bfseries 54} (2004) 1243}
  [\href{https://arxiv.org/abs/hep-th/0407187}{{\ttfamily hep-th/0407187}}].

\bibitem{glikman}
J.~Kowalski-Glikman, \emph{{Living in Curved Momentum Space}},
  \href{https://doi.org/10.1142/S0217751X13300147}{\emph{Int. J. Mod. Phys. A}
  {\bfseries 28} (2013) 1330014}
  [\href{https://arxiv.org/abs/1303.0195}{{\ttfamily 1303.0195}}].

\bibitem{juric2}
T.~Juri\'c, S.~Meljanac, D.~Pikuti\'c and R.~\v{S}trajn, \emph{{Toward the
  classification of differential calculi on \ensuremath{\kappa}-Minkowski space
  and related field theories}},
  \href{https://doi.org/10.1007/JHEP07(2015)055}{\emph{JHEP} {\bfseries 07}
  (2015) 055} [\href{https://arxiv.org/abs/1502.02972}{{\ttfamily
  1502.02972}}].

\bibitem{samsarov}
S.~Meljanac, A.~Samsarov, J.~Trampeti{\'c} and M.~Wohlgenannt, \emph{Scalar
  field propagation in the $\phi4\kappa$-minkowski model}, {\emph{JHEP}
  {\bfseries 2011} (2011) 1}.

\bibitem{dimitrijevic2}
M.~Dimitrijevic, L.~Jonke, L.~Moller, E.~Tsouchnika, J.~Wess and
  M.~Wohlgenannt, \emph{{Deformed field theory on kappa space-time}},
  \href{https://doi.org/10.1140/epjc/s2003-01309-y}{\emph{Eur. Phys. J. C}
  {\bfseries 31} (2003) 129}
  [\href{https://arxiv.org/abs/hep-th/0307149}{{\ttfamily hep-th/0307149}}].

\bibitem{meljanac1}
S.~Meljanac, A.~Pachol, A.~Samsarov and K.S.~Gupta, \emph{{Different
  realizations of \ensuremath{\kappa}-momentum space}},
  \href{https://doi.org/10.1103/PhysRevD.87.125009}{\emph{Phys. Rev. D}
  {\bfseries 87} (2013) 125009}
  [\href{https://arxiv.org/abs/1210.6814}{{\ttfamily 1210.6814}}].

\bibitem{lemos}
N.A.~Lemos, \emph{{Short proof of Jacobi's identity for Poisson brackets}},
  \href{https://doi.org/10.1119/1.19377}{\emph{Am. J. Phys.} {\bfseries 68}
  (2000) 88} [\href{https://arxiv.org/abs/physics/0210074}{{\ttfamily
  physics/0210074}}].

\bibitem{wohlgenannt}
S.~Meljanac, A.~Samsarov, J.~Trampetic and M.~Wohlgenannt, \emph{{Scalar field
  propagation in the $\phi^4$ kappa-Minkowski model}},
  \href{https://doi.org/10.1007/JHEP12(2011)010}{\emph{JHEP} {\bfseries 12}
  (2011) 010} [\href{https://arxiv.org/abs/1111.5553}{{\ttfamily 1111.5553}}].

\bibitem{kovacevic}
D.~Kovacevic and S.~Meljanac, \emph{{Kappa-Minkowski spacetime, Kappa-Poincare
  Hopf algebra and realizations}},
  \href{https://doi.org/10.1088/1751-8113/45/13/135208}{\emph{J. Phys. A}
  {\bfseries 45} (2012) 135208}
  [\href{https://arxiv.org/abs/1110.0944}{{\ttfamily 1110.0944}}].

\bibitem{rabin1}
R.~Banerjee, S.~Kulkarni and S.~Samanta, \emph{{Deformed symmetry in Snyder
  space and relativistic particle dynamics}},
  \href{https://doi.org/10.1088/1126-6708/2006/05/077}{\emph{JHEP} {\bfseries
  05} (2006) 077} [\href{https://arxiv.org/abs/hep-th/0602151}{{\ttfamily
  hep-th/0602151}}].

\bibitem{rabin2}
R.~Banerjee and S.~Samanta, \emph{{Gauge Symmetries on theta-Deformed Spaces}},
  \href{https://doi.org/10.1088/1126-6708/2007/02/046}{\emph{JHEP} {\bfseries
  02} (2007) 046} [\href{https://arxiv.org/abs/hep-th/0611249}{{\ttfamily
  hep-th/0611249}}].

\bibitem{hanson}
A.J.~Hanson, T.~Regge and C.~Teitelboim, \emph{{Constrained Hamiltonian
  Systems}}, Accademia Nazionale dei Lincei (1976).

\bibitem{carmona}
J.M.~Carmona, J.L.~Cort\'es and J.J.~Relancio, \emph{{Relativistic deformed
  kinematics from momentum space geometry}},
  \href{https://doi.org/10.1103/PhysRevD.100.104031}{\emph{Phys. Rev. D}
  {\bfseries 100} (2019) 104031}
  [\href{https://arxiv.org/abs/1907.12298}{{\ttfamily 1907.12298}}].

\bibitem{relancio}
M.~Arzano, G.~Gubitosi and J.J.~Relancio, \emph{{Deformed relativistic symmetry
  principles, arXiv:2211.11684}}, .

\bibitem{franchino}
S.A.~Franchino-Vi\~nas and J.J.~Relancio, \emph{{Geometrizing the
  Klein\textendash{}Gordon and Dirac equations in doubly special relativity}},
  \href{https://doi.org/10.1088/1361-6382/acb4d4}{\emph{Class. Quant. Grav.}
  {\bfseries 40} (2023) 054001}
  [\href{https://arxiv.org/abs/2203.12286}{{\ttfamily 2203.12286}}].

\bibitem{thanu}
T.~Padmanabhan, \emph{Gravitation Foundations and Frontiers}, Cambridge
  Monographs on Mathematical Physics, Cambridge University Press (2010),
  \href{https://doi.org/ISBN: 9780521882231}{ISBN: 9780521882231}.

\bibitem{arzano}
M.~Arzano and J.~Kowalski-Glikman, \emph{{Quantum particles in non-commutative
  space-time: an identity crisis, arXiv:2212.03703 [hep-th] (2022)}}, .

\bibitem{scholtz}
F.G.~Scholtz, B.~Chakraborty, S.~Gangopadhyay and A.G.~Hazra, \emph{{Dual
  families of non-commutative quantum systems}},
  \href{https://doi.org/10.1103/PhysRevD.71.085005}{\emph{Phys. Rev. D}
  {\bfseries 71} (2005) 085005}
  [\href{https://arxiv.org/abs/hep-th/0502143}{{\ttfamily hep-th/0502143}}].

\bibitem{stern}
L.~Lu and A.~Stern, \emph{Snyder space revisited},
  \href{https://doi.org/https://doi.org/10.1016/j.nuclphysb.2011.09.022}{\emph{Nuclear
  Physics B} {\bfseries 854} (2012) 894}.

\bibitem{pal}
S.K.~Pal and P.~Nandi, \emph{{Effect of dynamical noncommutativity on the
  limiting mass of white dwarfs}},
  \href{https://doi.org/10.1016/j.physletb.2019.134859}{\emph{Phys. Lett. B}
  {\bfseries 797} (2019) 134859}
  [\href{https://arxiv.org/abs/1908.11206}{{\ttfamily 1908.11206}}].

\bibitem{finsler}
Z.~Shen, \emph{Lectures on Finsler Geometry}, World Scientific (2001),
  \href{https://doi.org/https://doi.org/10.1142/4619}{https://doi.org/10.1142/4619}.

\bibitem{strassler}
M.J.~Strassler, \emph{{Field theory without Feynman diagrams: One loop
  effective actions}},
  \href{https://doi.org/10.1016/0550-3213(92)90098-V}{\emph{Nucl. Phys. B}
  {\bfseries 385} (1992) 145}
  [\href{https://arxiv.org/abs/hep-ph/9205205}{{\ttfamily hep-ph/9205205}}].

\end{thebibliography}\endgroup
\end{document}